\documentclass[prd, twocolumn, superscriptaddress, nofootinbib, longbibliography]{revtex4-2}

\usepackage{amsmath, amssymb}
\usepackage{graphicx}
\graphicspath{{figures_v8/}}
\usepackage[hidelinks]{hyperref}
\usepackage{booktabs}
\usepackage{bm}
\usepackage[dvipsnames]{xcolor}
\usepackage{enumitem}

\newcommand{\GeV}{\,\text{GeV}}
\newcommand{\MeV}{\,\text{MeV}}

\newcommand{\fbi}{\,\text{fb}^{-1}}
\newcommand{\fb}{\,\text{fb}}

\newcommand{\nb}{\,\text{nb}}

\newcommand{\he}{{}^3\text{He}}

\newcommand{\IA}{\text{IA}}

\newcommand{\pn}{pn}
\newcommand{\pp}{pp}
\begin{document}

\title{Isolating the axial two-body current via electromagnetic and
charged-current lepton scattering on polarized light nuclei}

\author{Guang Yang}
\affiliation{Brookhaven National Laboratory, Upton, NY 11973, USA}

\author{Praveen Kumar}
\affiliation{Brookhaven National Laboratory, Upton, NY 11973, USA}
\affiliation{University of Alabama, Tuscaloosa, AL 35487, USA}

\date{\today}

\begin{abstract}
Two-particle two-hole (2p2h) excitations driven by meson-exchange currents (MEC)
are among the leading nuclear uncertainties in long-baseline neutrino oscillation
experiments. The axial two-body current that drives the weak nuclear
responses is constrained by integrated weak rates at very low momentum transfer, tritium $\beta$-decay at $Q^2=0$ and muon capture on $^3$He at $Q^2\approx0.01\GeV^2$. We develop a method to isolate
that current. Electromagnetic scattering on a nucleus probes only the vector two-body current, while charged-current scattering on the same target probes the vector and axial currents together. Their difference isolates the axial-sensitive combination as a function of momentum transfer. Carried out on a sequence of polarized light nuclei, the method conditionally separates the $pn$ and $pp$ pair contributions to the vector and axial MEC. The spin-dependent response functions accessible with polarized beams decompose the
current into its underlying meson-exchange mechanisms. We quantify the
experimental requirements and find the program to be demanding: the dominant electromagnetic observables require an energy-transfer resolution of order $60\MeV$ and control of the impulse-approximation baseline at the few-percent level, while the charged-current measurement is limited by the separation of two forward neutrons and by data-driven control of background. 
If the axial strength is at least comparable to the vector one, a few hundred inverse femtobarns give a first direct test of the electromagnetic prediction.
The experimental requirements are not fully satisfied at present detectors but could lie within reach of a future high-luminosity polarized light-ion collider.
\end{abstract}

\maketitle

% =============================================================================
\section{Introduction}
\label{sec:intro}
% =============================================================================

The next generation of long-baseline neutrino experiments, such as DUNE~\cite{DUNEdesign} and Hyper-Kamiokande~\cite{HyperK}, aims for few-degree precision on the leptonic $CP$-violating phase $\delta_{CP}$ and a definitive determination of the neutrino mass ordering. Both experiments
infer the neutrino energy from the visible final state, assuming that the neutrino
struck a single, nearly stationary nucleon and the resulting nuclear smearing
is understood. That assumption carries large uncertainty
when the gauge boson couples to a correlated pair of nucleons through a
meson-exchange current, namely a two-particle two-hole (2p2h) process. Because such an
event leaves a muon and one or two undetectable nucleons in the detector, and is
indistinguishable from quasi-elastic scattering, the nucleons carry away
energy that is never recovered. Thus the reconstructed neutrino energy is biased low
by $50$ to $200\MeV$. The shift propagates directly into the extracted oscillation
parameters.

The 2p2h scattering accounting for the
MiniBooNE cross-section excess~\cite{MiniBooNE2010} was established by Martini and
collaborators~\cite{Martini2009,Martini2010} and independently by the Valencia
group~\cite{Nieves2011}. It has been confirmed across modern neutrino
data such as MINERvA~\cite{Rodrigues2016}. The difficulty is quantitative. The three 2p2h models in
common use in event generators, namely Valencia~\cite{Nieves2011}, SuSAv2-MEC~\cite{
Megias2016,SuSAv2GENIE}, and the Martini--Ericson model~\cite{MartiniGENIE2025}, differ at the tens-of-percent level for 2p2h. Recent generator benchmarks apply a representative $70\%$ normalization spread to cover the difference~\cite{MartiniGENIE2025,Benchmark2025,DUNE2p2h}. 

Electron scattering constrains part of the problem. The photon couples only to the
vector current, so electromagnetic data, including the recent JLab e4$\nu$
program~\cite{e4nu}, fix the vector two-body current but not directly the
axial one. The charged-current interaction relevant to neutrinos couples to
the sum of the vector and axial currents, and produces five independent nuclear
response functions~\cite{ODW1972,DonnellyPeccei1979,RuizSimo2017,Lovato2020},
\begin{equation}
\frac{d^2\sigma}{d\Omega\,d\omega} \propto
v_{CC} R^{CC} + v_{CL} R^{CL} + v_{LL} R^{LL}
+ v_{T} R^{T} + v_{T'} R^{T'}_{CC}.
\label{eq:five_responses}
\end{equation}
Among them, the transverse ($R^T$) and the charge and longitudinal responses
($R^{CC},R^{CL},R^{LL}$) have vector parts with direct electromagnetic analogs. The axial current contributes its $|J_A|^2$ terms to first four responses, whereas the pure vector--axial interference appears only in $R^{T'}_{CC}$. The two-body axial contributions have
never been measured directly. The existing constraints are two integrated weak rates at very low momentum transfer. The first is the Gamow--Teller matrix element
from tritium $\beta$-decay at $Q^2=0$~\cite{Gazit2009,Baroni2016}. The second is muon capture on $^3$He at $Q^2\approx0.01\GeV^2$~\cite{Ackerbauer1998,Marcucci2011}. The tritium point fixes a normalization and the capture rate tests it at small  momentum transfer. We refer to the absence at higher $Q^{2}$ as the axial gap.

This paper develops a way to address this gap. The central observation is that performing
both electromagnetic and charged-current lepton scattering on the same nuclear
target, and subtracting, isolates the axial-sensitive part of the two-body current.
Doing so on a sequence of light nuclei whose pair content is known, namely the proton, the
deuteron, $^3$He, and $^6$Li, separates the contributions of $pn$ and $pp$ pairs. In addition,
the polarized beams add the spin-dependent responses that distinguish the
underlying MEC mechanisms. A prior study analyzed the
charged-current process $e^-p\to\nu_e n$ on the free proton~\cite{YangKumar}. The identified background suppression is the controlling difficulty of that channel. That analysis provides the single-nucleon baseline.

Our emphasis is purely phenomenological. The measurements are not specific to
any accelerator, and the interesting question is not what a given facility would
achieve but what any apparatus must achieve to perform them. Taking that spirit, in this work, firstly, we present
the observables and their expected cross sections and yields, and then turn the
studies to quantify the required resolutions, acceptances, and luminosity, followed by the connection to the neutrino experiments.
The electromagnetic measurements are in part accessible
already at fixed-target electron facilities such as Jefferson Lab, where electron scattering in the dip region is an active program~\cite{e4nu} by running
polarized targets. The charged-current channel is
different. It needs collider energies to produce the $W$, and it needs the large
collider boost that sends the spectator nucleon forward where it can be tagged
precisely. 
These features define the instrument this program needs.
Following that, a polarized light-ion collider of the Electron-Ion Collider class~\cite{EICYellow}
is the natural setting as an example, being discussed and followed by the conclusion. 

% =============================================================================
\section{Full measurement program}
\label{sec:method}
% =============================================================================

The quantity we ultimately want is the strength of the two-body current as a
function of momentum transfer, separated into its vector and axial parts and into
the pair isospin channels. The method achieves it by measuring the cross-section excesses over
a single-nucleon baseline.
For each target and channel, we define a dimensionless excess ratio over the impulse
approximation (IA), with one nucleon at a time,
\begin{equation}
X(Q^2) \equiv \frac{\sigma_\text{meas}(Q^2)-\sigma_\IA(Q^2)}{\sigma_\IA(Q^2)},
\label{eq:excess}
\end{equation}
integrated over the dip region in energy transfer at fixed $Q^2$. Note that $\sigma_\IA$ is the full one-body response over the dip window. It combines quasi-elastic scattering with the single-pion continuum. Taking the excess against this full baseline sets the two-body enhancement at the $5$--$25\%$ level. 
%One consequence matters for the axial channel. The one-body $\Delta$ sits inside the charged-current baseline, so that baseline carries the axial $N\to\Delta$ form factor $C_5^A$.
%Within a single analysis this form factor is determined to better than ten percent~\cite{Graczyk2009}. The spread between analyses is larger, about $10$--$20\%$~\cite{Graczyk2009,Hernandez2010}. With the $\Delta$ covering roughly $40\%$ of the window, that adds a few-percent uncertainty to the charged-current baseline $\sigma_\IA^{\rm CC}$.

Two operational points fix the meaning of $\sigma_\IA$ and $\sigma_\text{meas}$. First, the excess of Eq.~(\ref{eq:excess}) is defined inclusively while we reconstruct the dip from exclusive final states. The struck nucleon with the tagged spectator in the electromagnetic channel and the two forward neutrons in the charged current are those final states. Only a one-body event has a clear spectator, and a two-body event ejects both nucleons into the same final state (Sec.~\ref{sec:fermisec}). The selection is quantified through the veto chain in Sec.~\ref{sec:mechreq}. Second, $\sigma_\IA$ is a plane-wave estimate, and final-state rescattering moves strength into the dip, where the one-body response is weakest, so a small rescattering effect can imitate a two-body excess. The low-spectator-momentum selection suppresses it, and the remainder is estimated from tagged-deuteron calculations in the quasi-elastic regime~\cite{BoeglinSargsian2015,CosynSargsian2011} and discussed in Sec.~\ref{sec:iabase}.

With the baseline fixed, the subtraction can be stated. Because the photon only couples to the vector current, the
electromagnetic excess measures the vector two-body strength alone. We denote it
$V_{ij}$, with $ij\in\{\pn,\pp\}$ labelling the pair. The charged-current excess on
the same target measures the vector and axial strengths together, $(V\!+\!A)_{ij}$.
%\revg{A word on notation, since it recurs throughout: $V$ and $A$ denote the vector and axial \emph{contributions to the measured cross section} (the rate), not
%the $V\!-\!A$ structure of the current operator. The weak charged current is indeed
%$V\!-\!A$; squaring it gives
%$|J_V-J_A|^2=|J_V|^2+|J_A|^2-2\,\mathrm{Re}(J_V J_A^*)$, so the vector %and axial
%strengths \emph{add} in the rate (each squared term is positive), together with a
%vector--axial interference term whose sign is process dependent. That additive
%structure is why the charged-current excess is written $(V\!+\!A)$.}
Their difference,
\begin{equation}
A_{ij}(Q^2) = (V\!+\!A)_{ij}(Q^2) - V_{ij}(Q^2),
\label{eq:axial}
\end{equation}
is the axial-sensitive combination.
Eq.~(\ref{eq:axial}) retains both the pure axial term $|J_A|^2$ and the vector--axial
interference $-2\,\mathrm{Re}(J_V J_A^*)$. The interference sits in the part of the cross section that flips sign with the beam helicity. That part scales with the inelasticity $y$, and at a collider, $y$ is only about $10^{-4}$. Therefore, its weight is only a few $\times10^{-4}$ of the unpolarized
charged-current rate. So the measured $A_{ij}$ is the essentially pure axial two-body
strength $|J_A|^2$. We use $|J_A|^2$ as shorthand for the operationally defined axial excess.

The subtraction assumes that the vector two-body current measured electromagnetically equals the vector part of its charged-current counterpart. This is the conserved vector current (CVC) hypothesis, where the weak vector current and the isovector electromagnetic current are two components of one conserved isospin current. This assumption is suitable here because the dominating mechanisms in the dip window are all isovectors. 

Two details follow when the CVC hypothesis written out. First, the two currents differ by a fixed Clebsch--Gordan factor. It cancels in the excess ratio of Eq.~(\ref{eq:excess}), where it multiplies the numerator and the denominator similarly. Second, the charged-current baseline $\sigma_\IA^{\rm CC}$ carries the one-body axial response on top of the vector one. We have to compute that axial part from the measured nucleon axial form factor. Hold that separate, both ratios refer to the same one-body vector response. 

The only thing that the CVC does not transport is the isoscalar part of the electromagnetic current. For two-body currents, it is small and comes mostly from $\rho\pi\gamma$ exchange~\cite{Riska1989}. The elastic deuteron observables bound it independently, with the clearest signature in $A(Q^2)$ at large $Q^2$~\cite{GilmanGross} (Sec.~\ref{sec:em}).
%The direct charged-current measurement supplies the cross-check.

It should be noted that the subtraction does not require the two channels to be equal in absolute size. The electromagnetic and charged-current cross sections differ by several orders of
magnitude, but the fractional two-body enhancement is comparable in the two channels, in the
order of $5$--$25\%$.
Since the two excess ratios are of the same size, the subtraction in
Eq.~(\ref{eq:axial}) is between two comparable numbers, even though the underlying cross sections are not.

Getting to this point so far, the subtraction still just concerns a single target. Separating the pair types is what the sequence of targets for. The four pair-separated currents, namely $V_{\pn}$, $V_{\pp}$, and their
charged-current partners, are disentangled by the known pair content of the targets.
The free proton has no pair and carries no
two-body excess. It calibrates the single-nucleon form factors that fix the impulse-approximation baseline $\sigma_\IA$ common to every target. The deuteron is a single $pn$ pair, so its dip excess gives $V_{\pn}$ in the electromagnetic channel and $(V\!+\!A)_{\pn}$ in the charged-current channel directly, with no
isospin ambiguity. Helium-3 holds two $pn$ pairs and one $pp$ pair, so its excess
is a density-weighted sum of the two pair types,
\begin{equation}
X(\he) = f_{\pn}\,V_{\pn} + f_{\pp}\,V_{\pp},
\label{eq:isospin}
\end{equation}
where $f_{\pn}$ and $f_{\pp}$ are calculable pair-density factors from the $A=3$
wave function. Subtracting the deuteron isolates $V_{\pp}$, and the
identical method in the charged-current channel gives $(V\!+\!A)_{\pp}$. 

The per-pair subtraction in Eq.~(\ref{eq:axial}) is exact for the $pn$ pair under CVC, since both probes reach the same isovector matrix elements. On the other hand, for the $pp$ pair, the seagull and pion-in-flight currents carry the isospin factor $(\bm\tau_1\times\bm\tau_2)$, whose charge-conserving component annihilates a $pp$ state. Two protons cannot exchange a charged pion and remain two protons. The electromagnetic probe at $T_z=+1$ does not see this strength  while the charged current does. Then the difference $(V\!+\!A)_{\pp}-V_{\pp}$ retains uncancelled vector strength and is not a clean axial observable. Therefore, we quote the directly measured $(V\!+\!A)_{\pp}$ as the $pp$ deliverable. Importing the missing matrix elements from the $pn$ sector by isospin symmetry is possible in principle and is left for future work.

Nevertheless, one gain appears based on the same selection rule. With the seagull and pion currents switched off, the electromagnetic $pp$ excess isolates the $\Delta$ mechanism, leading to the mechanism decomposition for Measurement~III introduced later. 
%It is the most uncertain step because the $pp$ extraction is a difference of two large quantities and the model dependence on the $A=3$ ground state and
%the density factors. Thus 
The $pp$ extraction is also developed separately in Appendix~\ref{app:pp}. 
%The deuteron $pn$ results are more direct.

\begin{figure*}[t]
\centering
\includegraphics[width=\textwidth]{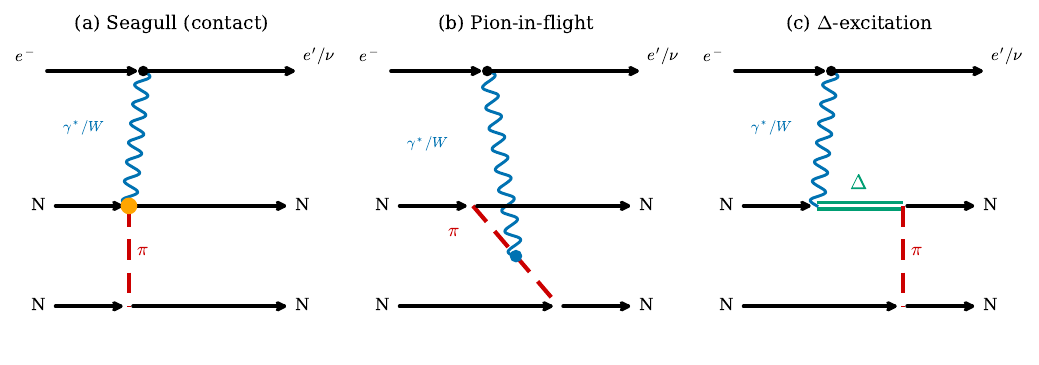}
\caption{The three meson-exchange mechanisms that dominate the dip, drawn for a
leptonic probe that exchanges a $\gamma^*$ (electromagnetic) or a $W$ (charged
current) with a nucleon pair: (a) the contact seagull current, (b) the
pion-in-flight current, and (c) $\Delta$-isobar excitation. A fourth mechanism,
$\rho\pi\gamma$ heavy-meson exchange, is subdominant in the dip and is not shown
(Tab.~\ref{tab:mech}). 
The star on $\gamma^*$ marks a virtual ($Q^2>0$)
photon exchanged in the scattering, as distinct from a real on-shell photon
($Q^2=0$). }
\label{fig:feynman}
\end{figure*}

Now up to this point, the two-body current has been treated as one number per pair and channel. It also has internal structure and that structure is measurable. The two-body current includes four mechanisms~\cite{Riska1989}.
Three dominate the dip, shown in Fig.~\ref{fig:feynman}, including a contact (seagull) current, in which the boson couples to the pion--nucleon vertex, a pion-in-flight current, in which it couples to the exchanged pion, and $\Delta$-isobar excitation, in which a nucleon is promoted to a $\Delta(1232)$ that de-excites by pion exchange.
A fourth, the $\rho\pi\gamma$ heavy-meson exchange current, is numerically smaller
in the dip thus omitted in the decomposition framework. Even though, its elastic signature and the $T_{20}$ test of the two-body charge operators are discussed in Sec.~\ref{sec:em}.

The mechanisms carry distinct spin--isospin and orbital structure, summarized in Tab.~\ref{tab:mech}.
All of them are isovector, except for the $\rho\pi\gamma$
current, which is purely isoscalar by G-parity~\cite{MRS1998,GilmanGross,CarlsonSchiavilla1998}. Its isovector partner is the separate $\omega\pi\gamma$ current, excluded from the CVC map and bounded by isovector observables. 
The $\rho\pi\gamma$ charge operator matters for the elastic $A(Q^2)$ at large $Q^2$ while its transverse piece in the dip is small. The two roles are kept distinct in Sec.~\ref{sec:em}. The $\Delta$ current is the only one that is purely spin-flip
($\Delta S=1$), giving a unique signature in the tensor responses.
The seagull and pion-in-flight currents carry spin-flip pieces as well, but mixed with non-flip strength.
Each mechanism contributes to a given response function with a calculable weight $w_m^\alpha$. 

For a polarized response function $\Delta R^\alpha$
measured relative to the vector strength,
\begin{equation}
\frac{\Delta R^\alpha(Q^2)}{V_{\pn}(Q^2)} =
\sum_{m\in\{\text{sea},\pi,\Delta\}} w_m^\alpha\, f_m(Q^2),
\qquad \sum_m f_m = 1,
\label{eq:weights}
\end{equation}
with $f_m$ the fraction of the two-body strength carried by mechanism $m$. These fractions are defined within a given current model. The model also fixes how the interference between mechanisms is shared among them. So the fractions sum to one by construction.

%The dip region for our signal is dominated by the three mechanisms shown in Fig.~\ref{fig:feynman}. The $\rho\pi\gamma$ contribution enters mainly
%through the elastic deuteron form factors at large $Q^2$. The elastic observables carry two-body corrections at the ten-percent level~\cite{SchiavillaRiska1991,CarlsonSchiavilla1998}. In order to make the paper efficient and effective, the $\rho\pi\gamma$ term is omitted from the fractions in Eq.~(\ref{eq:weights}). Its growth with $Q^2$ is bounded by the elastic constraint, and the residual it leaves in the dip fit is carried as a systematic toward $Q^2\approx1\GeV^2$.
%Because the eight deuteron response functions~\cite{DonnellyRaskin,Arenhovel2005} weight the three mechanisms differently, they form an overdetermined linear system for the two independent fractions.
%The seagull and pion-in-flight currents share the isospin and orbital pattern of Tab.~\ref{tab:mech}. At fixed $Q^2$ they are separated by their different weights across the responses, and across $Q^2$ by their different trends.

The full program determines which microscopic
current dominates the two-body strength as a function of $Q^2$. The largest difference among the neutrino interaction models is their mechanism content. The Valencia and SuSAv2 calculations differ mainly in
the weight they assign to $\Delta$ excitation, so a direct determination of the fractions discriminates among them at the level of the underlying physics. The cleanest handle is the tensor-transverse
response, for which the $\Delta$ weight is negative while the seagull and
pion-in-flight weights are positive. So a $\Delta$-dominated current produces a sign
change in the dip. Binding effects are not expected to mimic it and establishing this requires the dedicated calculation flagged in Sec.~\ref{sec:mechreq}.

The mechanism decomposition also connects the vector and axial currents, allowing the statistics-sufficient electromagnetic data to constrain the axial gap even if we are facing a difficult charged-current measurement. Each mechanism carries a ratio of its axial to vector strength, $(A/V)_m$, estimated within a given two-body-current model. In the current models these ratios are in the order of one. We adopt $(A/V)\sim1$ as the working expectation below. The axial
two-body current also has a pion-pole (induced-pseudoscalar) part with no vector
counterpart~\cite{BaroniAxial2016}, so the $(A/V)_m$ carry genuine model uncertainty. That is why the direct charged-current subtraction is needed to test the prediction. 

Once the fractions $f_m$ are known from the electromagnetic responses, the axial-sensitive excess is predicted from
the measured vector excess as
\begin{equation}
A_{\pn}^\text{pred}(Q^2) = \Big[\textstyle\sum_m (A/V)_m\, f_m(Q^2)\Big]\,V_{\pn}(Q^2).
\label{eq:vaconnection}
\end{equation}
The direct charged-current subtraction of Eq.~(\ref{eq:axial}) then becomes a test
of this prediction. The agreement validates the operator picture and a discrepancy
would point to new axial dynamics. The two routes to $A_{\pn}$, the
indirect one through Eq.~(\ref{eq:vaconnection}) and the direct one through
Eq.~(\ref{eq:axial}), set different demands on the detector.

\begin{table}[t]
\caption{Spin--isospin structure of the four meson-exchange mechanisms.
IV (IS) denotes isovector (isoscalar). The ``$L$ channels'' row lists the
relative orbital angular momenta of the nucleon pair. That ``specific'' means that the $\Delta$
current, a pure spin-flip $N\to\Delta$ transition, is restricted to particular
spin-triplet channels rather than the general $S/D$ mixture. }
\label{tab:mech}
\begin{ruledtabular}
\begin{tabular}{lcccc}
& Seagull & $\pi$-flight & $\Delta$ & $\rho\pi\gamma$ \\
\colrule
Isospin & IV & IV & IV & IS \\
Spin-flip & partial & partial & pure & partial \\
$L$ channels & $0,2$ & $0,2$ & specific & $0,2$ \\
$Q^2$ trend & flat & $1/(Q^2{+}m_\pi^2)$ & peaks $0.3$--$1$ & grows \\
\end{tabular}
\end{ruledtabular}
\end{table}

Furthermore, a parity-violating asymmetry offers an independent, electromagnetic handle on
the axial response. The small left--right asymmetry from $\gamma$--$Z$ interference
in polarized electron scattering is in principle sensitive to the nuclear axial two-body current without any charged-current measurement. The required sub-ppm control of the asymmetry has been demonstrated on the proton by the Qweak measurement~\cite{Qweak2018}. That sensitivity enters through the electron's vector coupling $g_V^e=-1+4\sin^2\theta_W\approx-0.08$, accidentally small relative to the axial coupling $g_A^e$, so the axial-current piece is a few parts-per-million effect sitting on the $\sim\!25$ ppm asymmetry (the latter set by $g_A^e$ times the nuclear weak charge). 
%It is dominated by helicity-correlated systematics rather than statistics. 
This asymmetry offers a cross-check on the subtraction shown in
Eq.~(\ref{eq:axial}). On the other hand, because the axial term is carried by the beam-helicity
asymmetry, it shares the small-inelasticity suppression of the other primed observables
(Sec.~\ref{sec:helicity}), resulting in the fact that it is more suitable to be a fixed-target measurement.

Overall, the program breaks into five measurements, developed in the following sections.
\begin{itemize}[leftmargin=1.4em,itemsep=3pt,topsep=2pt,label={\textendash}]
\item Measurement~I (vector responses, deuteron): Unpolarized and
polarized electron--deuteron scattering in the dip yields the vector excess
$V_{\pn}$ and the spin-dependent response functions, on a pure $pn$ pair.
The unpolarized vector excess and the tensor-polarized responses are collider
observables. The double-spin (beam\,$\times$\,vector-target) responses $R^{T'}_{\rm EM}$ and $R^{TL'}$ are kinematically suppressed at collider kinematics, thus more suitable as fixed-target measurement on deuteron and $^3$He (Sec.~\ref{sec:helicity}).
\item Measurement~II (axial isolation): Charged-current scattering on the
deuteron and on $^3$He gives $(V\!+\!A)_{\pn}$. Both targets probe the same $pn$ axial current. They are
complementary such that the deuteron is isospin-pure $pn$ and $^3$He carries a
taggable proton spectator.
The CC$-$EM subtraction of Eq.~(\ref{eq:axial}) then returns the axial-sensitive $A_{\pn}(Q^2)$.
A vector-polarized $^3$He target adds the target-spin asymmetry $A_T^{CC}$ as a two-body observable.
\item Measurement~III (mechanism decomposition): The tensor
responses, accessible with a tensor-polarized deuteron (the interference responses only at fixed target, Sec.~\ref{sec:helicity}), separate the seagull,
pion-in-flight, and $\Delta$-excitation contributions and predict $A_{\pn}$ from the
vector data alone.
\item Measurement~IV ($pp$ pairs): Adding $^3$He, and $^6$Li for the
tensor sector, extends the decomposition to $pp$ pairs (Appendix~\ref{app:pp}).
\item Measurement~V (cross-checks): The elastic deuteron tensor
analyzing power and a parity-violating asymmetry test the same exchange-current framework independently from the dip program.
The elastic $T_{20}$ uses a tensor-polarized target as a collider observable.
The parity-violating axial term, being beam-helicity-correlated, is a fixed-target measurement (Sec.~\ref{sec:helicity}).
\end{itemize}

% =============================================================================
\section{The electromagnetic measurements (MEAS. I and III)}
\label{sec:em}
% =============================================================================

The electromagnetic measurements on the deuteron anchor measurements that follow.
The deuteron is the cleanest realization of the method. It is a single loosely
bound $pn$ pair, so any excess over the single-nucleon baseline is the $pn$
two-body current with no isospin ambiguity. In addition, the unstruck nucleon recoils as a
single spectator that can be detected to fix the initial state. The electromagnetic
measurements on the deuteron are statistics-rich and form the robust baseline measurement of the full program.

The relevant kinematics are those located at the 2p2h dip.
As the energy transfer $\omega$
is increased at fixed $Q^2$, the inclusive response shows a quasi-elastic peak near
$\omega_\text{QE}=Q^2/2M_N$, a $\Delta$ resonance peak near
$\omega_\Delta=(M_\Delta^2-M_N^2+Q^2)/2M_N$, and a dip between them where both
single-nucleon processes are falling and the relative two-body contribution is
largest.
For example, at $Q^2=0.5\GeV^2$ the window is bounded by the two peak positions,
$\omega_\text{QE}=0.27\GeV$ and $\omega_\Delta=0.61\GeV$, the dip being the valley between them. We integrate observables over this window
and present them as functions of $Q^2$. For all quoted excesses and yields this
peak-to-peak window is narrowed to the guarded signal region of Sec.~\ref{sec:rates}, which
excludes a guard of $\approx110\MeV$ around each peak. The quasi-elastic peak is not a sharp line but a band. The struck nucleon is already moving inside the deuteron when the probe arrives, and that motion smears the peak. The size of this internal momentum follows from the weak binding. With a binding energy $B=2.2\MeV$ and a nucleon mass $M_N$, it is $p_F\approx\sqrt{M_N B}\approx45\MeV$. The peak then acquires a width $\sim\!|\bm q|\,p_F/M_N\approx40\MeV$, where $\bm q$ is the three-momentum transfer. This intrinsic width, along with the detector resolution,
is critical to the requirements of Sec.~\ref{sec:req}.

The deuteron is a spin-1 target, and its inclusive polarized cross section
separates into eight independent response functions, in contrast to two
for an unpolarized spin-0 or spin-1/2 target, and four for a spin-1/2
target with polarized beam and target~\cite{DonnellyRaskin,Arenhovel2005}.
Four tensor responses are unique to spin $\geq1$. With an unpolarized beam there are the longitudinal
$R^L$, coupling to the nuclear charge, and the transverse $R^T$, coupling to the current. Since meson-exchange currents are currents, they enhance $R^T$ by tens of percent in the dip, leaving $R^L$ almost unchanged, so the transverse excess is the dominant vector signal and defines $V_{\pn}\equiv
\Delta R^T/R^T_\IA$. 

One practical point follows. The signal is transverse, but the measurement cannot isolate $R^T$ by itself. At $\varepsilon\approx1$ the cross section responds to the sum $R^T+R^L$ (Sec.~\ref{sec:ltsep}), so what the data return is $X=\Delta R^T/(R^T_\IA+R^L_\IA)$. The numerator is unaffected, since the longitudinal excess stays negligible. Only the baseline in the denominator has grown. Recovering $V_{\pn}$ therefore needs no further measurement, just the one-body baseline the method already requires, through $V_{\pn}=X\,(1+R^L_\IA/R^T_\IA)$. That factor is $1.2$ to $1.5$ in the dip at $Q^2=0.5\GeV^2$. We quote following results in the $V_{\pn}$ convention.
The charged-current excess is converted to the same convention before the subtraction, as described in Sec.~\ref{sec:cc}.

\begin{figure*}[t]
\centering
\includegraphics[width=\textwidth]{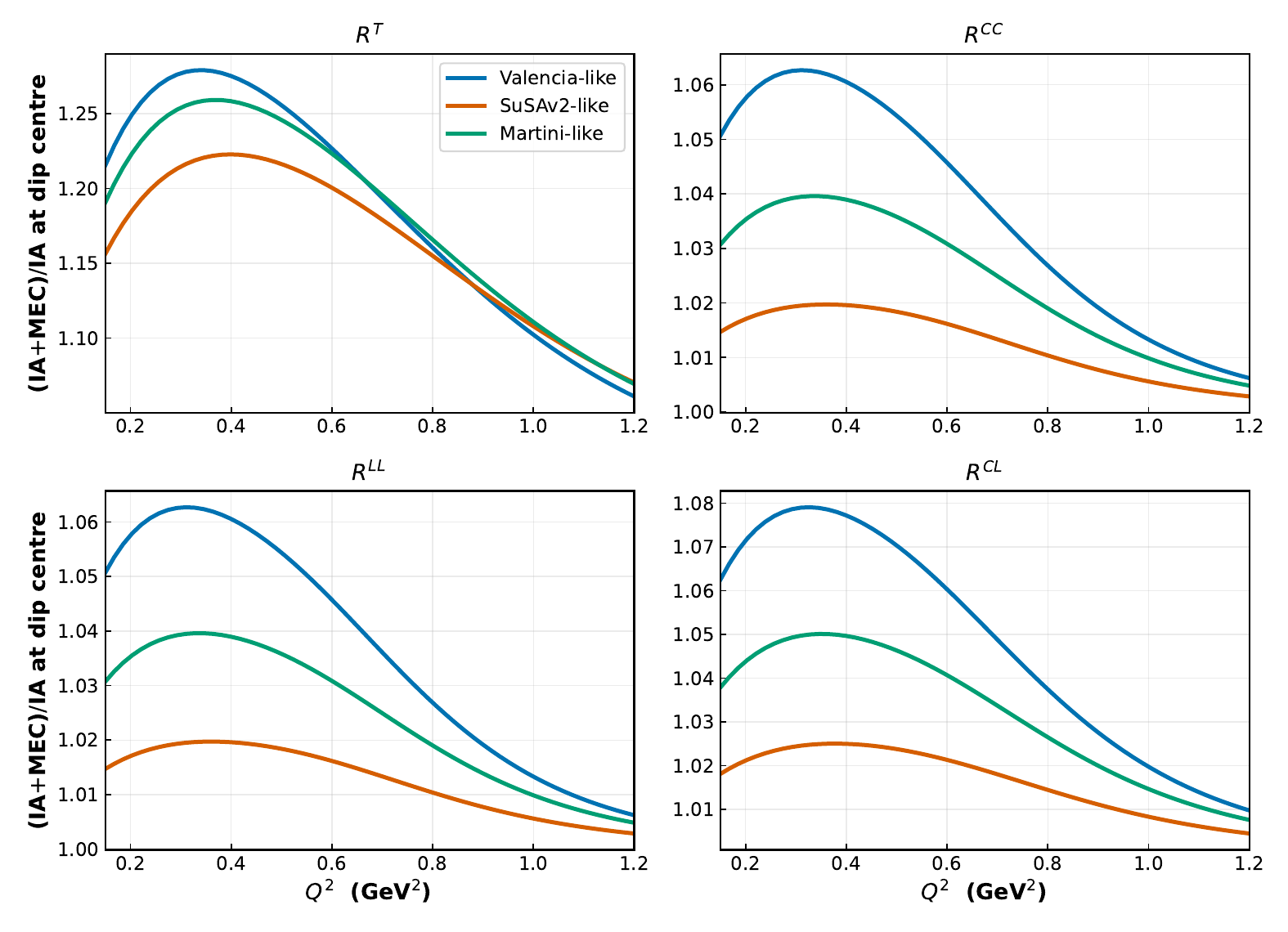}
\caption{Expected two-body enhancement of the dip response functions as functions of $Q^2$, shown per channel as the ratio of the full response to the one-body baseline, both evaluated at the dip centre $\omega^*(Q^2)$. The one-body baseline at that point is the quasi-elastic tail plus the one-body $\Delta$. The local enhancement at the valley floor goes high while the excess integrated over the guarded window is the $15$ to $25\%$ quoted elsewhere. The curves are illustrative parameterizations spanning the range of published model properties~\cite{Nieves2011,Megias2016,MartiniGENIE2025}.
The four panels are the transverse response $R^T$, the Coulomb $R^{CC}$, the longitudinal $R^{LL}$, and the charge--longitudinal interference $R^{CL}$.}
\label{fig:responses}
\end{figure*}

A polarized beam on a vector-polarized deuteron adds the interference responses $R^{T'}_{\rm EM}$ and $R^{TL'}$ through the double-spin asymmetry (suppressed at collider kinematics, Sec.~\ref{sec:helicity}). In addition, a tensor-polarized deuteron adds the four tensor responses $R^L_{20}$, $R^T_{20}$,
$R^{TT}_{22}$, and $R^{TL}_{21}$ through the analyzing power and the azimuthal
modulation of the cross section.
Here the tensor analyzing power is the relative difference in cross section between
the two tensor-polarization states of the deuteron (the magnetic substate
$m\!=\!0$ versus $m\!=\!\pm1$). On the other hand, the azimuthal modulation is the dependence of the inclusive cross section on the orientation of the target polarization (the deuteron spin direction) relative to the lepton scattering plane. Both are accessible only with a polarized spin-1 target and isolate the tensor responses.
The four tensor responses vanish identically for
spin-0 and spin-1/2 targets. They arise from the quadrupole (D-state) component of the deuteron wave function and its interference with the S-state. The tensor-transverse response $R^T_{20}$ carries the $\Delta$ sign test mentioned in Sec.~\ref{sec:method}. Most of these responses have never been measured on any target in the dip region. Tab.~\ref{tab:obs} lists the eight responses with the polarization each requires, as well as
the information each carries. In Fig.~\ref{fig:responses}, we show the expected two-body enhancement of the response functions, as
predicted by the three model-like parameterizations over the one-body baseline. It sets the scale of the signal in the measurement.

\begin{table}[t]
\caption{The eight deuteron response functions, the polarization each requires, and
what each isolates. $P_e$, $P_z$, and $P_{zz}$ denote beam, vector-target, and
tensor-target polarization. The four tensor responses ($L_{20}$, $T_{20}$,
$TT_{22}$, $TL_{21}$) are unique to a spin-1 target. Rows are grouped by facility:
the beam-helicity (double-spin) responses $R^{T'}_{\rm EM}$ and $R^{TL'}$ are suppressed at
collider kinematics and better suited to fixed-target experiments, while the unpolarized and tensor-polarized
responses are collider observables (Sec.~\ref{sec:helicity}).}
\label{tab:obs}
\begin{ruledtabular}
\begin{tabular}{lll}
Response & Requires & Measures \\
\colrule
\multicolumn{3}{c}{\emph{Collider observables}} \\
$R^L$ & unpolarized & charge response (IA) \\
$R^T $ & unpolarized & vector $pn$ current \\
$R^L_{20}$ & $P_{zz}$ & D-state $+$ charge MEC \\
$R^T_{20}$ & $P_{zz}$ & $\Delta$ sign test \\
$R^{TT}_{22}$ & $P_{zz}$ & quadrupole / mechanism \\
$R^{TL}_{21}$ & $P_{zz}$ & $L$--$T$ tensor interference \\
\colrule
\multicolumn{3}{c}{\emph{Fixed-target observables}} \\
$R^{T'}_{\rm EM}$ & $P_e,P_z$ & spin-flip interference \\
$R^{TL'}$ & $P_e,P_z$ & $L$--$T$ spin-flip \\
\end{tabular}
\end{ruledtabular}
\end{table}

Before turning to the charged-current channel, we would like to describe an independent
cross-check that reaches the same two-body physics through a different reaction.
Everything above concerns the inelastic dip, in which the deuteron breaks up.
The deuteron can instead be probed elastically, staying intact and
recoiling as a whole. Elastic
electron--deuteron scattering is described by three form factors: the charge
monopole $G_C$, the charge quadrupole $G_Q$, and the magnetic dipole $G_M$~\cite{GilmanGross}.
Because the deuteron is a spin-1 system with a D-state mixture, $G_C$ is controlled mainly
by the S-state and $G_Q$ by the D-state. The observable that
isolates their interplay is the tensor analyzing power $T_{20}$, the fractional
change in the elastic cross section when the target tensor polarization is switched between its $m=0$ and $m=\pm1$ substates. The $T_{20}$ is dominated by the $G_C G_Q$ interference and it measures the S--D overlap directly. Note that the $T_{20}$ is large and negative in this region, reaching its kinematic limit $-\sqrt{2}$ below the $G_C$ node near $Q^2\approx0.7\GeV^2$~\cite{GilmanGross}.
That has been mapped precisely by BLAST~\cite{BLAST_T20},
Novosibirsk~\cite{NIKHEF_T20}, and JLab~\cite{JLab_T20}. The connection to this work
is that two-body charge operators, together with relativistic corrections, move
the zero position and change the magnitude of $T_{20}$ at the ten-percent level~\cite{SchiavillaRiska1991,CarlsonSchiavilla1998}. The $\rho\pi\gamma$ current itself is predominantly transverse and its elastic signature is carried by $A(Q^2)$ at large $Q^2$~\cite{GilmanGross}. The existing data already confirm that the two-body currents are present with expected size. Extending the measurement to a few $\GeV^2$, where the models
diverge without data, would turn this consistency check into a quantitative test of the same exchange-current framework that fills the dip.

% =============================================================================
\section{The charged-current measurement (MEAS. II)}
\label{sec:cc}
% =============================================================================

The axial information requires the charged-current channel. The interaction final state matters for the experiment sensitivities. A $W^-$ converts a proton
to a neutron, so charged-current quasi-elastic scattering on the deuteron,
$e^-d\to\nu_e\,nn$, produces two neutrons without a proton. The advantage the
deuteron has in the electromagnetic channel, that a charged spectator proton is well measured, is missing in the charged-current channel. The spectator is a neutron, which is caught in a forward hadronic
calorimeter. The background is then harder to control. 
%Even though, the deuteron remains valuable as the isospin-pure $pn$ reference. %but may not be the cleanest charged-current target.

Furthermore, this final state limitation is intrinsic. On the deuteron the quasi-elastic and two-body channels end in the same two-neutron final state without charged tracks, although the reconstructed $\omega^*$ itself is safe, due to the fact that summing over both measured neutrons cancels the internal momentum (Sec.~\ref{sec:fermisec}).  
On the contrary, with a clearly tagged slow spectator, the struck nucleon carries low initial momentum. Such a quasi-free sample anchors the baseline (Sec.~\ref{sec:iabase}). Therefore, the deuteron charged-current channel serves as the isospin-pure reference while the baseline anchoring lives on $^3$He.
The $^3$He restores the proton tagging advantage. It has three nucleons, two protons and one
neutron ($ppn$). A charged-current event acts on one $pn$ pair, converting it to two neutrons, $W^-+pn\to nn$. The third nucleon, the other proton, is untouched and recoils as a spectator. That spectator proton can be tagged precisely, resulting in the fact that $^3$He provides the better-constrained initial state for the axial measurement. 

However, the $W^-$ can also act on the $pp$ pair, turning it into $np$ with the neutron as the untouched spectator. That topology also ends in two neutrons and a proton, so neutron counting alone does not separate the two pair types. Instead, the proton momentum does. In the $pn$ case the tagged proton is a slow spectator carrying only Fermi momentum. In the $pp$ case the proton is a struck nucleon and comes out fast. Requiring a slow tagged proton selects the $pn$ pair, and the tagged $^3$He excess measures $(V\!+\!A)_{\pn}$.  

The two targets complement each other such that the deuteron
fixes the $pn$ channel cleanly in isospin, and $^3$He supplies the well-tagged
charged-current measurement. Combining the charged-current excess on $^3$He with the vector excess from the deuteron through Eq.~(\ref{eq:axial}) yields the
axial-sensitive $A_{\pn}(Q^2)$. Before the subtraction, the charged-current excess is converted to the $V_{\pn}$ convention shown in Sec.~\ref{sec:em}, with a factor of roughly $1.1$ to $1.3$ computed from the same one-body baseline. The factor is smaller than the electromagnetic one because the one-body axial strength is predominantly transverse. Its small uncertainty is carried in the baseline budget. 

One systematic attaches to this cross-nucleus subtraction is that the $pn$ pair in $^3$He is denser than the loosely bound deuteron pair, so the per-pair excess is not identical on the two targets. We carry this density mismatch in the baseline budget alongside the $C_5^A$ term. 
%The all-deuteron subtraction is free of the systematic. 
On the other hand, it lacks a spectator tag so its $\omega^*$ carries the full Fermi smearing described above. Therefore, it serves as a consistency check at coarser resolution. Given
the charged-current channel is limited by both backgrounds and counting~\cite{YangKumar},
it returns
a single integrated axial-sensitive excess instead of a separation of the
individual weak response functions in Eq.~(\ref{eq:five_responses}). The
spin-dependent separation is the province of the statistics-rich electromagnetic channel. This direct measurement tests the indirect prediction in Eq.~(\ref{eq:vaconnection}). In other words, the electromagnetic mechanism decomposition forecasts $A_{\pn}$ from the vector excess, and the charged-current measurement confirms it.

Another practical issue is the $Q^2$ assignment on the charged-current side. A charged-current event has no scattered charged lepton so its $Q^2$ must come from the hadronic side. Transverse-momentum conservation gives $Q^2_{\rm rec}=|\bm p_{T,n1}+\bm p_{T,n2}|^2$, equating $Q^2$ to the accuracy of the small inelasticity. With the $80\MeV$ pair resolution (derived in Sec.~\ref{sec:ccdip}), the relative resolution is $2\,\delta q_T/q_T\approx23\%$ at $Q^2=0.5\GeV^2$, against bin half-widths of $15\%$. A simulation of the migration gives the diagonal fractions in Tab.~\ref{tab:migration}.
Subtracting a sharp electron-side $V_{\pn}$ from this smeared measurement bin by bin would bias the shape even at infinite statistics. We can do forward-fold. The measured $V_{\pn}(Q^2)$ is smeared with the same migration response and the subtraction of Eq.~(\ref{eq:axial}) is performed in reconstructed $Q^2$. The migration response also enters the missing-transverse-momentum selection and the two are varied together in the systematic budget.

{\color{blue}\begin{table}[t]
\caption{Diagonal fractions of the charged-current $Q^2$ migration for the dip bins with the $80\MeV$ pair resolution. }
\label{tab:migration}
\begin{ruledtabular}
\begin{tabular}{lc}
True bin ($\GeV^2$) & Same-bin fraction \\
\colrule
$0.20$--$0.35$ & $57\%$ \\
$0.35$--$0.50$ & $49\%$ \\
$0.50$--$0.65$ & $44\%$ \\
$0.65$--$0.80$ & $40\%$ \\
$0.80$--$1.00$ & $46\%$ \\
\end{tabular}
\end{ruledtabular}
\end{table}}

Furthermore, a polarized target adds one further charged-current observable with no
electromagnetic counterpart. The charged-current target-spin asymmetry $A_T^{CC}$ is
formed by flipping the target polarization while collecting charged-current events. Both this asymmetry and the response $R^{T'}_{CC}$ of Eq.~(\ref{eq:five_responses}) come from vector--axial interference, so it is worth keeping them apart. 
The response $R^{T'}_{CC}$ is the interference term of the unpolarized cross section. It is the piece that makes neutrino and antineutrino rates differ and it needs no polarized target, given that the $W$ already picks out a single lepton helicity. 

The asymmetry $A_T^{CC}$ is a different quantity. It appears only when the target is polarized, and it measures how the two-body current couples to the spin of the struck pair.
On $^3$He the one-body part of this asymmetry is small. Almost all of the $^3$He spin sits on the single neutron~\cite{Friar1990}. An $e^-$ beam delivers a $W^-$ that turns a proton into a neutron. The probe strikes the unpolarized protons rather than the polarized neutron. A one-body event then shows little target-spin asymmetry. What survives comes mainly from the two-body currents, making $A_T^{CC}$ unique.

Two points follow. The first is statistics. The rate gives $\approx15$ events per $Q^2$ bin per $50\fbi$. Even a few hundred $\fbi$ leave a per-bin error of $\approx10\%$. The asymmetry is also diluted, because only the two-body part of the rate carries it, and that part is $15$ to $25\%$ of the total. Therefore, we treat $A_T^{CC}$ as a high-luminosity goal rather than a primary observable.

The second is that the argument above holds for $^3$He alone. A vector-polarized deuteron behaves differently. There the struck proton carries part of the target spin itself. The one-body asymmetry is no longer suppressed, and it dominates the measurement. The two targets thus probe different quantities. Only $^3$He gives the clean two-body signal.

In addition, the extension to $pp$ pairs requires subtracting the scaled
deuteron from $^3$He. For the $pp$ tensor responses, a spin-1 target such as $^6$Li is needed. This part is more model-dependent and described in Appendix~\ref{app:pp}.
In Tab.~\ref{tab:inventory}, we summarize the full set of observables across the targets
and channels, measurement requirement, and the quantity it constrains. The
four primary unknowns are the measured pair excesses
$V_{\pn}$, $V_{\pp}$, $A_{\pn}$, and $(V\!+\!A)_{\pp}$, the axial isolation being available for the $pn$ pair only. The mechanism fractions add two more independent parameters.

\begin{table*}[t]
\caption{Complete observables. EM = electromagnetic ($\gamma^*$),
CC = charged current ($W^-$), PV = parity-violating.}
\label{tab:inventory}
\begin{ruledtabular}
\begin{tabular}{lllll}
Observable & Channel / pol. & Target & Constrains & Facility \\
\colrule
$\Delta R^L,\ \Delta R^T\ (V_{\pn})$ & EM & $d$ & vector $pn$ current; IA check & collider \\
$\Delta R^{T'}_{\rm EM},\ \Delta R^{TL'}$ & EM, $P_e P_z$ & $d$ & $J_V$ spin structure & fixed target \\
$\Delta R^L_{20},\Delta R^T_{20},\Delta R^{TT}_{22},\Delta R^{TL}_{21}$ & EM, $P_{zz}$ & $d$ & mechanism fractions; $\Delta$ sign & collider \\
$T_{20}$ (elastic) & EM, $P_{zz}$ & $d$ & two-body charge operators; framework check & collider \\
$(V\!+\!A)_{\pn}$ & CC & $d$ & isospin-pure $pn$ vector$+$axial & collider \\
$A_T^{CC}$ & CC, $P_z$ & $\he$ & two-body pair-spin & collider \\
$\Delta R^{L},\ \Delta R^{T}$ & EM & $\he$ & $V_{\pn}\!+\!V_{\pp}$ & collider \\
$\Delta R^{T'}_{\rm EM}$ & EM, $P_e P_z$ & $\he$ & $J_V$ spin structure & fixed target \\
$(V\!+\!A)_{\pn}$ (slow-$p$ tag) & CC & $\he$ & isospin-selected $pn$ vector$+$axial & collider \\
$(V\!+\!A)$ (inclusive) & CC & $\he$ & $(V\!+\!A)_{\pn}\!+\!(V\!+\!A)_{\pp}$ & collider \\
$\delta A_{\rm PV}$ & PV, $P_e$ & $d,\he$ & axial cross-check & fixed target \\
\colrule
\multicolumn{5}{c}{\textit{Derived}} \\
$A_{\pn}=(V\!+\!A)_{\pn}-V_{\pn}$ & CC$-$EM & $\he$/$d$ & axial-sensitive $pn$ (principal) & \\
$V_{\pp},\,(V\!+\!A)_{\pp}$ & $\he-$scaled $d$ & $\he$ & $pp$ pair currents, axial isolation $pn$ only & \\
$f_\text{sea},f_\pi,f_\Delta$ & weight matrix & $d$ & mechanism content & \\
$A_{\pn}^\text{pred}$ & V--A connection & $d$ & test of direct $A_{\pn}$ & \\
\end{tabular}
\end{ruledtabular}
\end{table*}

% =============================================================================
\section{Expected cross sections and yields}
\label{sec:rates}
% =============================================================================

The absolute scales that drive the requirements are set by the well-understood quasi-elastic cross sections. We quote rates per $Q^2$ bin of width $0.15\GeV^2$. Within a bin, we integrate over a central dip signal region in $\omega^*$ that excludes a region of $\approx110\MeV$ around the quasi-elastic and $\Delta$ peaks. 

The region's width is chosen so that peak leakage into the signal region stays at the few-percent level for the $60\MeV$ resolution. At $Q^2=0.5\GeV^2$, the quasi-elastic electromagnetic cross section is in the order of $40\nb$ per bin, and over the central dip region the one-body response is at the nb scale. These are standard form-factor results. The model dependence is carried by the $\sim\!20\%$ two-body enhancement on top of them. Ab-initio Green's-function Monte Carlo (GFMC) calculations~\cite{Lovato2020} and the phenomenological 2p2h models such as Valencia~\cite{Nieves2011} and SuSAv2~\cite{Megias2016} all predict a substantial enhancement but differ in its size and composition.
Their spread over this enhancement sets the precision a measurement must
reach (Sec.~\ref{sec:req}).

The corresponding charged-current scale is the quasi-elastic cross section on the free proton, $\sigma_\text{CC}\approx9\fb$ at these energies, computed from the Llewellyn Smith formalism~\cite{LlewellynSmith1972} with the measured vector and axial form factors. This number anchors the charged-current yields below and the background estimate shown in Sec.~\ref{sec:bkg}. It is worth noting that this estimate corresponds to  left-handed electron beam. Every yield scales with $(1+P_e)/2$, a $\approx10\%$ correction at the $P_e\approx0.8$ of a collider beam.

Overall, the electromagnetic dip channel delivers
an order of $10^{6}$--$10^{7}$ events per $Q^2$ bin per $\fbi$, so that counting statistics are far
below the two-body signal at modest integrated luminosity. Instead, the
electromagnetic program is limited by the impulse-approximation baseline and the dip resolution (Sec.~\ref{sec:req}). 

The charged-current channel is opposite. For example, the interaction on $^3$He is estimated from the $9\fb$ scale through the following chain, with the two-neutron acceptance carried per bin. The two active protons on $^3$He double the rate. The dip range $Q^2=0.2$--$1.0\GeV^2$ holds about half of the total, and the bin at $Q^2=0.5\GeV^2$ carries roughly a fifth of that dip range yield. The signal region holds a quarter to a third of the strength in the bin, counting the window share of the one-body strength anchored to the quasi-elastic bin yield. 

Requiring both forward neutrons inside the zero-degree acceptance is a strong, $Q^2$-dependent selection. The pair centroid sits at $\sqrt{Q^2}/(2\gamma m_N)$ off axis and each neutron adds an excursion of $p^*/\gamma m_N$ around it, so both angles grow with $Q^2$ while the aperture is fixed. As shown in Fig.~\ref{fig:acc2n}, a toy simulation of the pair decay gives the acceptance $\varepsilon_{2n}$ as a band spanning the $\omega^*$ sampling choices within the dip. At maximum deuteron rigidity ($\gamma\approx146$) with a $5.5$~mrad neutral aperture, $\varepsilon_{2n}$ is close to unity at $Q^2=0.2\GeV^2$, between $0.5$ and $0.8$ at $0.5\GeV^2$, and between $0.15$ and $0.2$ at $1.0\GeV^2$. At $\gamma=100$, with a $4$~mrad aperture, the acceptance goes below tens of percent already at $0.2\GeV^2$ and collapses below the percent level over the rest of the dip. Therefore, the charged-current program requires the maximum-rigidity configuration and the enlarged aperture. All charged-current yields below carry $\varepsilon_{2n}(Q^2)$.

{\color{blue}\begin{figure}[t]
\centering
\includegraphics[width=\columnwidth]{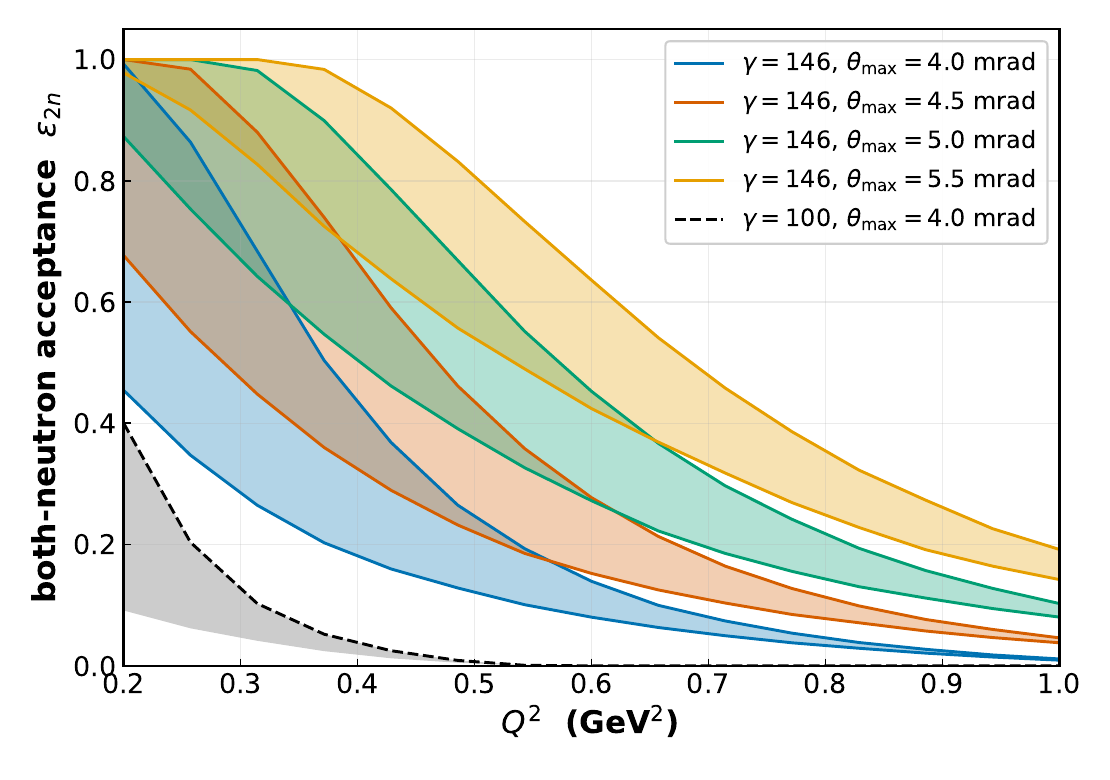}
\caption{Both-neutron zero-degree acceptance $\varepsilon_{2n}$ versus $Q^2$ from a toy simulation of the $nn$ pair decay, for four apertures at maximum deuteron rigidity and for the $\gamma=100$, $4$~mrad configuration. Each band spans the $\omega^*$ sampling choices within the dip, from the guarded signal window to a central band, with isotropic decay in the pair frame. A MEC-weighted angular distribution will be added as a further variation.}
\label{fig:acc2n}
\end{figure}}

Putting all numbers together, the lowest bins hold $20$ to $30$ events per bin per $50\fbi$ where $\varepsilon_{2n}$ approaches unity. The central bins hold $12$ to $24$ events with the acceptance band. The highest bin falls to $3$ to $6$ events. These counts do not yet include the per-neutron detection and two-cluster reconstruction efficiencies, which enter squared and appear as separate requirement rows in Tab.~\ref{tab:req}. That said, the axial measurement is counting-limited, and thus sets the luminosity requirement. We quote yields per unit luminosity throughout and do not assume a particular run plan. The free proton's electromagnetic cross section is known to high precision, providing an in-situ calibration of the luminosity and acceptance to all targets. %It also sets a charged-current reference through $e^-p\to\nu_e n$~\cite{YangKumar}.

% =============================================================================
\section{Experimental requirements}
\label{sec:req}
% =============================================================================

With the information above, now we look into the apparatus requirement for each step of the program. 
The requirements can be categorized into three groups in the order of increasing difficulty.
The first, shared by the electromagnetic measurements, covers resolving the dip
(Sec.~\ref{sec:dipres}), the longitudinal/transverse separation
(Sec.~\ref{sec:ltsep}), the suppression of helicity-dependent observables at collider kinematics (Sec.~\ref{sec:helicity}), the Fermi-motion correction (Sec.~\ref{sec:fermisec}), and
the impulse-approximation baseline (Sec.~\ref{sec:iabase}). The second is the beam
polarization that the mechanism decomposition needs (Sec.~\ref{sec:mechreq}). The
third is the charged-current channel (Secs.~\ref{sec:ccdip} and \ref{sec:bkg}) that contains the hardest demands. The
conclusions are collected in Tab.~\ref{tab:req}.

\subsection{Resolving the dip}
\label{sec:dipres}

The two-body signal lives in the dip, a window $\sim\!0.34\GeV$ wide in the
energy transfer $\omega^*$ measured in the nucleon rest frame. A finite resolution
$\sigma_\omega$ smears the quasi-elastic and $\Delta$ peaks into the valley. The
fraction of a peak that leaks past a boundary a distance R away is
$\tfrac12\,\mathrm{erfc}(R/\sqrt2\,\sigma_\omega)$. The relevant distance is from a peak to the edge of the signal region, e.g. $R\approx110\MeV$ at $Q^2=0.5\GeV^2$. Keeping that leakage below a few percent of a peak then requires $\sigma_\omega\lesssim60\MeV$. A few percent of the quasi-elastic peak is comparable to the two-body signal itself. The leaked strength is a smooth and calculable smearing of the one-body peaks, and it folds into the baseline $\sigma_\IA$. What the measurement needs is that the uncertainty on this leakage stays below the few-percent baseline budget in Sec.~\ref{sec:iabase}. 

\begin{figure*}[t]
\centering
\includegraphics[width=0.8\textwidth]{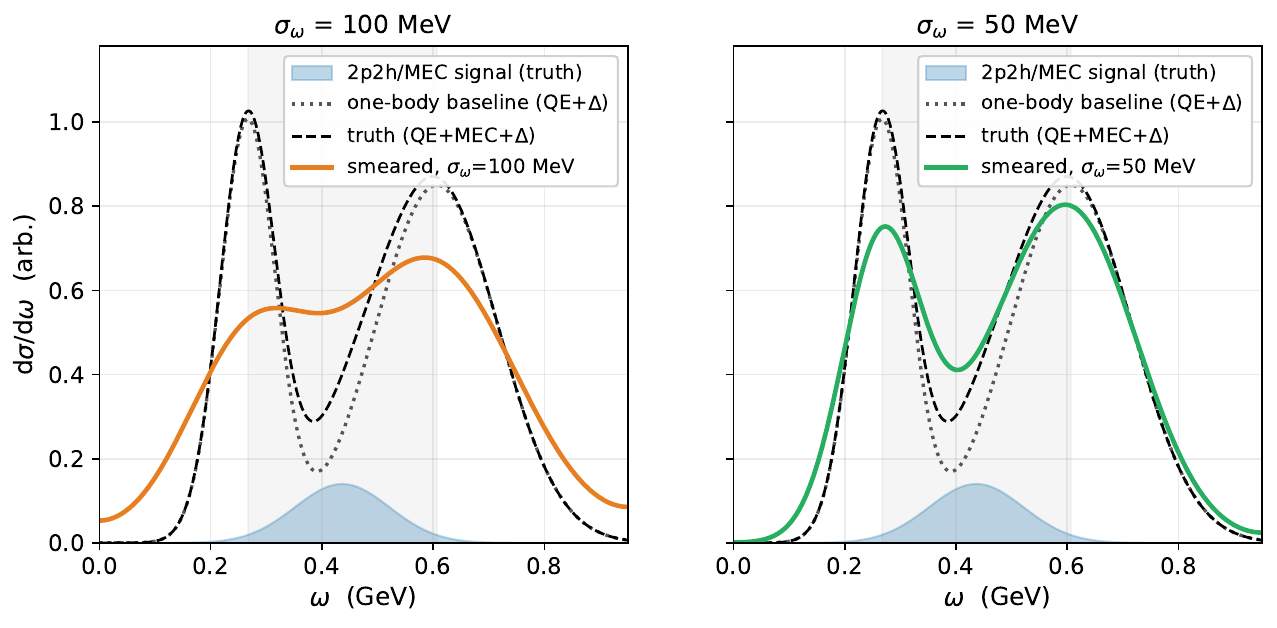}
\caption{Illustrative dip spectrum in the nucleon rest frame, smeared by an
energy-transfer resolution $\sigma_\omega$. The dip survives at $\sigma_\omega=
50\MeV$ but is washed out by $100\MeV$. The dotted curve is the one-body
$\sigma_\IA$ baseline (quasi-elastic plus the one-body $\Delta$). The shaded band
is the 2p2h/MEC signal. }
\label{fig:smear}
\end{figure*}

In Fig.~\ref{fig:smear}, we show
an illustrative dip spectrum, built from the quasi-elastic peak, the
$\Delta$ peak, and the small two-body bump, after smearing by two values of $\sigma_\omega$. At $50\MeV$ the dip is still clearly resolved, whereas at $100\MeV$ the two large peaks leak across the valley.
This is a stringent but not
extreme requirement on the rest-frame reconstruction.

The way this resolution is achieved matters. In fixed-target experiments, one
might expect to take $\omega^*$ from the scattered electron. At collider kinematics this fails. The struck nucleon is itself a fast
beam, the inelasticity $y$ is in the order of $10^{-4}$, and then the rest-frame $\omega^*$ inferred from the scattered electron carries an uncertainty of tens of GeV due to the boost.
The dip must instead be reconstructed from the hadronic final state together with the spectator tag, or from the full nucleon pair when both are struck (Sec.~\ref{sec:fermisec}).

In the electromagnetic channel, above methodology is within reach. When the struck nucleon is a proton, it is tracked. The combination $E-p_z$ of its energy and longitudinal
momentum (the ``light-cone'' momentum, which is the quantity least distorted by the
boost) fixes $\omega^*$ to well within the requirement.
The struck proton emerges fast and almost along the beam, so its energy $E$ and longitudinal
momentum $p_z$ are individually large and each is strongly distorted by the
boost. Their difference $E-p_z$ is instead small (order of $m_N^2/E$). Under a
boost along the beam it simply scales by an overall factor, leaving its
fractional resolution unchanged. Because the rest-frame energy transfer
$\omega^*$ is carried by this small light-cone component, measuring $E-p_z$
determines $\omega^*$ robustly. 

Reconstructing $\omega^*$ from the hadrons brings one more benefit. An electron can radiate a photon before or after the scattering. The radiated energy is lost from the electron side, so a reconstruction based on the scattered electron places such an event at the wrong $\omega^*$. In inclusive experiments this smears the large quasi-elastic peak into a tail that leaks into the dip, at the level of several to tens of percent of the local one-body strength. Here the hadronic final state records the energy actually delivered to the nucleons. A radiated quasi-elastic event therefore is recorded at its true $\omega^*$ and does not fake the dip strength. What remains is a small migration in $Q^2$ that carried by the electron, and the usual radiative correction to the free-proton calibration.

The difficulty
is specific to the charged-current channel, where a $W^-$ turns the struck proton
into a neutron and the recoil can only be measured in a forward calorimeter.
The time-of-flight can hardly supply the energy given these forward neutrons are ultrarelativistic. The background is correspondingly harder. Reaching the requirement is a genuine detector challenge, demanding future high-granularity forward-neutron detection technology developments such as~\cite{Agarwal2023,Gwon2023,oWbLS2026}, and specifically the high-granularity zero-degree calorimetry being developed~\cite{PaulZDC2024}.

\subsection{The longitudinal/transverse separation}
\label{sec:ltsep}

A clean energy-transfer reconstruction is necessary but not sufficient. The electromagnetic measurement meets another purely kinematic limit when one tries to separate its two unpolarized responses. Separating $R^L$ from $R^T$ requires
measuring the cross section at several values of the virtual-photon polarization $\varepsilon$, varied by changing the beam energy. 
The polarization $\varepsilon$
runs from $0$ (back-scattering) to $1$ (forward scattering), and the reduced cross section is linear in it, $\sigma_\text{red}=R^T+\varepsilon R^L$. 

At collider kinematics, the electron barely deflects, so $\varepsilon$ is pinned near unity at every available energy. On the other hand, the relevant $\varepsilon$ is the rest-frame one as the
response functions $R^L$ and $R^T$ are defined in the nucleon rest frame. Because the target is ultrarelativistic
($\gamma\approx100$), the electron energy in that frame is several thousand GeV and the
scattering angle is sub-milliradian. So $\varepsilon$ moves from unity by only
$\sim\!10^{-8}$--$10^{-7}$ across the three beam energies (Tab.~\ref{tab:epsilon}).
That means the lever arm is $\Delta\varepsilon\sim3\times10^{-7}$ and thus a Rosenbluth $L/T$
separation is impossible. 
We quote $Q^2=0.5\GeV^2$ as a
representative dip-center point, but this conclusion holds across the whole
dip-accessible range, $Q^2\approx0.2$--$1.0\GeV^2$.

Although a clean longitudinal/transverse separation in the dip is not feasible, this does
not compromise the vector measurement because the two-body excess is
almost purely transverse. At $\varepsilon\approx1$ one measures the sum $R^T+R^L$ rather than the two separately. For the vector current, conservation ties the longitudinal response to the charge density and limits the
longitudinal piece of the excess to an estimated $\lesssim\!5\%$~\cite{Lovato2016}.
Therefore, the electromagnetic excess is $\Delta R^T$ to that accuracy. 

On the other hand, the axial current does not obey such conservation law. Its charge and longitudinal components survive at full kinematic weight in the charged-current channel. 
%Only the pion-pole part  is removed by the electron mass. Here, the pion pole is the induced-pseudoscalar piece of the axial current, in which the exchanged boson converts to a pion carrying the four-momentum transfer proportional to $q^\mu$. Contracting $q^\mu$ with the lepton current yields a factor of the lepton mass, negligible for an electron beam. 
This axial-longitudinal strength cannot be separated out by the measurement due to the aforementioned kinematics suppression. 
So we have to treat it as part of the measured axial excess. In the program, the direct subtraction includes it by construction while the indirect prediction built from the transverse mechanisms does not. That mismatch is carried as an explicit term in the comparison of the two routes, to be bounded by a dedicated evaluation of the five weak responses at dip kinematics (Sec.~\ref{sec:mechreq}).
%What got lost is the explicit $\Delta R^L$ cross-check. The residual few-percent longitudinal contamination is carried as a systematic on $V_{\pn}$.
\begin{table}[t]
\caption{Virtual-photon polarization $\varepsilon$ in the nucleon rest frame at
$Q^2=0.5\GeV^2$ (dip center) for the three electron beam energies, with the
rest-frame electron scattering angle $\theta_e^*$.}
\label{tab:epsilon}
\begin{ruledtabular}
\begin{tabular}{ccc}
$E_e$ (GeV) & $\theta_e^*$ (mrad) & $1-\varepsilon$ \\
\colrule
18 & $0.20$ & $2.7\times10^{-8}$ \\
10 & $0.35$ & $8.7\times10^{-8}$ \\
5  & $0.71$ & $3.5\times10^{-7}$ \\
\end{tabular}
\end{ruledtabular}
\end{table}

\subsection{Helicity-dependent observables at collider kinematics}
\label{sec:helicity}

The same kinematics that closes the $L/T$ separation also decides which spin
observables survive. Every response that couples to the beam helicity correlated
(antisymmetric) part of the lepton tensor, the primed responses, carries a kinematic weight
proportional to $\tan(\theta_e^*/2)$. With the sub-milliradian rest-frame angles shown in
Tab.~\ref{tab:epsilon}, this weight is tiny. To be more specific, evaluating the standard kinematic factors at the
paper's own kinematics gives
\begin{equation}
\frac{v_{T'}}{v_T} = (2\text{--}8)\times10^{-4}
\label{eq:vtprime}
\end{equation}
across the three beam energies at $Q^2=0.5\GeV^2$, against $v_{T'}/v_T\approx0.4$ at the JLab fixed-target point ($2.2\GeV$, same $Q^2$). Here $v_T$ and $v_{T'}$ are the kinematic weights that multiply $R^T$ and $R^{T'}$ in the cross section as shown in Eq.~(\ref{eq:five_responses}). Their ratio sets the largest helicity-dependent response. This sorts the program between the two kinds of facility.

First, the double-spin electromagnetic responses $R^{T'}_{\rm EM}$ and $R^{TL'}$ (the
polarized part of Measurement~I) are not accessible at the collider. The double-spin
asymmetry amplitude is bounded by $v_{T'}/v_T\sim$~few~$\times10^{-4}$, so even $10^7$ events per bin give barely a $1\sigma$ measurement of the asymmetry itself. That argument makes these measurements fixed-target observables. Second, the parity-violating axial cross-check (Measurement~V)
carries the analogous factor $\varepsilon'=\sqrt{\tau(1+\tau)}\sqrt{1-\varepsilon^2}
\approx10^{-4}$ in the rest frame (versus $\approx0.16$ at the JLab point). So the few-ppm
axial piece in Sec.~\ref{sec:method} is also a fixed-target measurement. And third, the
vector--axial interference in the unpolarized charged-current cross section carries the
same suppression so it contributes only a few $\times10^{-4}$ of the rate. As anticipated in
Eq.~(\ref{eq:axial}), this reduces the measured $A_{ij}$ to the pure axial strength $|J_A|^2$.

Bulk of the program still stays at the collider. The unpolarized vector excess, the four tensor responses, and the charged-current yield all couple to the symmetric part of the lepton tensor. The weights of that part, $v_L$, $v_T$, $v_{TL}$, and $v_{TT}$, stay finite as $\theta_e^*\to0$. So none of these observables is suppressed. In addition, the tensor modulations need an azimuthal reference. The lab-frame electron provides it at $\theta_e^{\rm lab}\approx40$~mrad, far from the disappearing rest-frame angle.

On the other hand, the scattered electron carries requirements as well. The electron defines $Q^2$ and it supplies the azimuthal reference. Because $y$ is tiny, the electron keeps nearly the full beam energy and emerges at $\theta_e^{\rm lab}\approx\sqrt{Q^2}/E_e$. At $E_e=18\GeV$, the dip range $Q^2=0.2$--$1.0\GeV^2$ sits at $25$ to $56$~mrad, respect to pseudorapidity $3.6$ to $4.4$. At $E_e=5$ to $10\GeV$, the same range sits at $45$ to $200$~mrad. That means the electron-side acceptance favors the lower beam energies. Furthermore, since $\varepsilon$ is pinned at unity for all three settings, the helicity arguments are unchanged by that choice.

As a special case, the target-spin asymmetry $A_T^{CC}$ also survives due to parity. A single target-spin asymmetry is a parity-odd observable. The electromagnetic current conserves parity. In there the asymmetry can only be built with the help of the beam helicity while that carries the $\tan(\theta_e^*/2)$ suppression. The weak current violates parity on its own. Its target-spin interference enters through a spin structure function of the $g_5$ type, which rides on the symmetric part of the lepton tensor. At the kinematics shown in Tab.~\ref{tab:epsilon}, those weights are about order of one against the beam-helicity weight $10^{-4}$. So $A_T^{CC}$ stands as a collider observable.

Overall, the vector polarization of the deuteron beam was needed only for the double-spin responses. Then the collider beam request reduces to tensor polarization alone. The mechanism decomposition is unchanged. It remains overdetermined by the surviving responses $R^T$, $R^T_{20}$, $R^{TT}_{22}$, and $R^L_{20}$, and the $\Delta$ sign test sits in $R^T_{20}$ (Sec.~\ref{sec:mechreq}).

\subsection{Fermi motion and spectator tagging}
\label{sec:fermisec}

Resolving the dip and separating the responses both presume knowledge of the
struck nucleon's initial motion. A
nucleon bound in the deuteron carries an internal (Fermi) momentum in the order of $45\MeV$, set by the binding ($\sqrt{M_N B}$) with a tail to higher values. Boosted
into the collider frame with $\gamma\approx100$, this becomes a longitudinal
ambiguity of order several $\GeV$. This boost would make
an exclusive reconstruction impossible. The remedy is to tag the spectator. In the rest frame, the
struck nucleon's initial momentum is fixed by the spectator,
$\bm p_\text{struck}=-\bm p_\text{spectator}$, and a forward tracking detector
measuring a charged spectator to $\delta p/p\sim10^{-3}$ determines it well. 

The rest-frame spectator momentum is reconstructed component by component, and
the two components behave very differently. The longitudinal component is compressed by the inverse boost, so that the large lab momentum divided by
$\gamma$ leaves a measurement error of only $\sim\!1\MeV$. The transverse component is
not compressed. It is limited by the beam angular divergence times the beam momentum,
$\sim\!15\MeV$. This dominates the net resolution. The tag therefore fixes the
initial nucleon momentum to $\sim\!15\MeV$, against an intrinsic Fermi spread of
$\sim\!45\MeV$. 

\begin{figure}[t]
\centering
\includegraphics[width=\columnwidth]{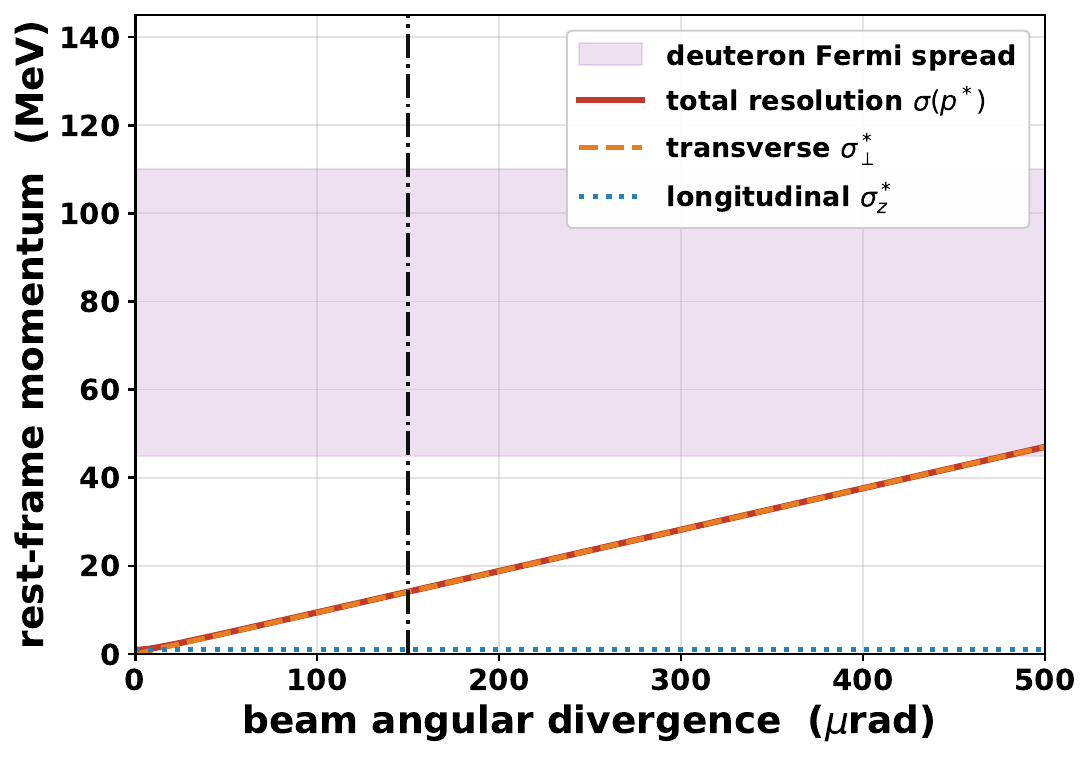}
\caption{Resolution on the rest-frame spectator (Fermi) momentum as a function
of the beam angular divergence for a far-forward spectator tagging
($\delta p/p\sim10^{-3}$). The dash-dotted line marks the nominal EIC divergence ($\sim\!150\,\mu$rad),
where the total resolution ($\sim\!15\MeV$) lies well below the intrinsic deuteron
Fermi spread (band). }
\label{fig:fermi}
\end{figure}

In Fig.~\ref{fig:fermi}, we show how the rest-frame spectator resolution depends on the beam angular divergence. The
longitudinal component is set by the tracker momentum resolution divided by the boost
and stays near $1\MeV$,
independent of the divergence. At the nominal EIC divergence
of $\sim\!150\,\mu$rad, the total resolution is about $15\MeV$. It remains below the intrinsic deuteron Fermi spread as long as the divergence stays under
$\sim\!480\,\mu$rad.

This argument applies to the electromagnetic channel and the charged-current channel on $^3$He when the
spectator is a proton. 
On the deuteron, the proton spectator arises when the photon strikes the neutron, roughly a fifth of the rate at these $Q^2$. When the photon strikes the proton the spectator is a neutron, measured only in the zero-degree calorimeter at coarser resolution. The statistics in Sec.~\ref{sec:rates} leave the struck-neutron sample ample for the precision baseline program.

The spectator does more than correct the kinematics. Binning events in the spectator momentum selects different regions of the nuclear wave function. Low
momentum ($|\bm p_s|\lesssim100\MeV$) picks the quasi-free S-wave configuration,
where the impulse-approximation baseline is most reliable (Sec.~\ref{sec:iabase}).
High momentum ($|\bm p_s|\gtrsim300\MeV$) reaches the short-range correlated region,
where the two-body currents are expected to be largest~\cite{SRC}.
The charge of the spectator also
identifies which nucleon was struck, separating $pn$ from $pp$ knockout in the
heavier targets. Both capabilities require forward acceptance over a wide spectator-momentum range.

Then note a two-body event may eject both nucleons. Aforementioned tagging serves the baseline more than the signal. 
Two-body events populate the region with both nucleons are fast. The excess is measured from the full dip yield against the baseline extrapolated from the quasi-free sample. The $\omega^*$ reconstruction is unaffected because the full $pn$ final state is measured in both event classes. 

The Fermi motion does not degrade the two-body events either.  
There are two points to make it stand. 
The first is the pair total is known given the sum is a property of the beam and not of the wave function. The internal momentum does not enter the kinematics.  The beam energy is defined to a few parts in $10^4$, which moves $\omega^*$ by well under an MeV. Meanwhile, the beam direction is defined to the $\sim\!150\,\mu$rad divergence, which enters through the transverse momentum transfer at the $\sim\!10\MeV$ level. The second is the boost cancellation. Each nucleon separately carries the boosted internal spread of several GeV of longitudinal momentum. The sum removes the spread exactly, then the detector errors enter only through $E-p_z$. For the single zero-degree neutron, the smearing is at the $\approx\!30\MeV$ level. So the reconstruction is limited by the neutron calorimeter.

\subsection{The impulse-approximation baseline}
\label{sec:iabase}

\begin{figure}[t]
\centering
\includegraphics[width=\columnwidth]{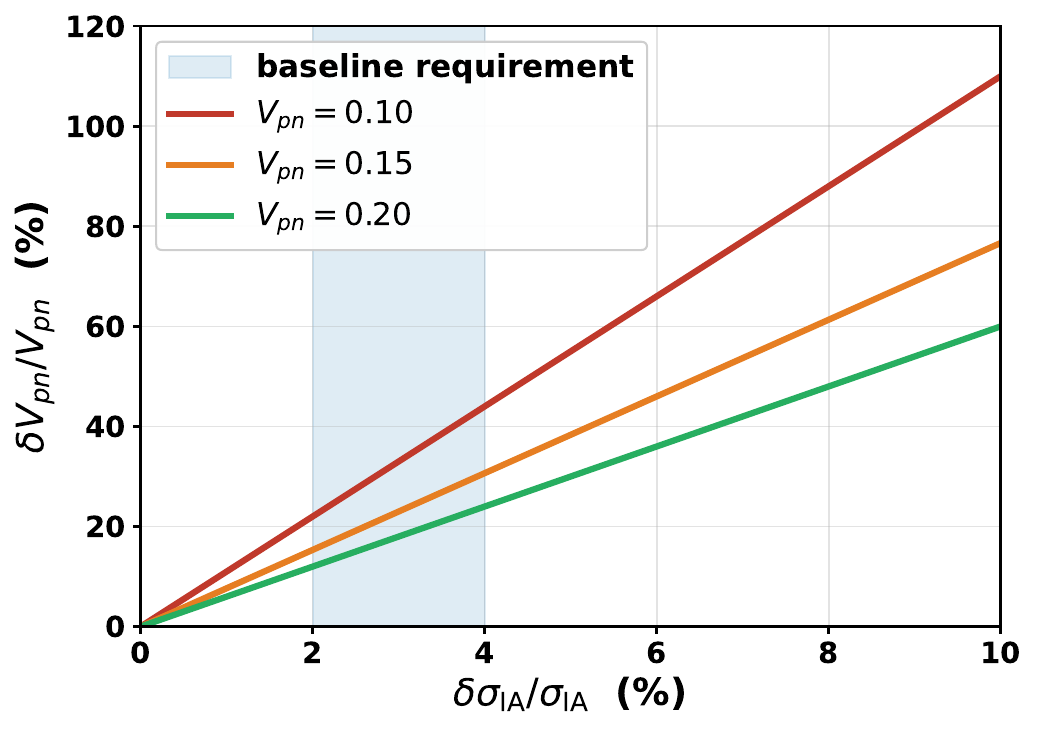}
\caption{Achievable relative precision on the vector excess $V_{\pn}$ as a
function of the impulse-approximation baseline uncertainty
$\delta\sigma_\IA/\sigma_\IA$, for three values of the signal $V_{\pn}$. Separating
the leading models requires the baseline known to the few-percent level. The shaded band marks the $2$--$4\%$ requirement of Tab.~\ref{tab:req}.}
\label{fig:ia}
\end{figure}

The signal sits on a baseline
that must be known independently of the detector. Every observable is an excess at the order of $20\%$ over the impulse-approximation baseline, so the baseline itself must be
controlled. Propagating its uncertainty $r$
gives $\delta V_{\pn}/V_{\pn}\approx(1+V_{\pn})r/V_{\pn}$, so resolving the
tens-of-percent model spread requires $r$ at the few-percent level. In Fig.~\ref{fig:ia}, this propagation for several signal sizes is shown. 
In the low-spectator-momentum sample where the baseline control is actually achievable, the excess is smaller than the inclusive $20\%$. So the relevant curves in Fig.~\ref{fig:ia} are with $V_{\pn}=0.10$ to $0.15$ while the requirement on $r$ tightens toward $2$ to $3\%$.

The baseline uncertainty must reach
the few-percent level for the achievable precision on $V_{\pn}$ to be small enough to
separate the models. Whether this is achievable is controlled by the spectator
momentum. Selecting low spectator momentum picks the quasi-free regime, where the
deuteron wave function is precise~\cite{AV18} and final-state interactions are small. There the
vector (electromagnetic) baseline is known to $1$--$3\%$. 

The main concern is the rest of the one-body yield, fed by larger initial momenta with more uncertainty on the wave function. The signal window maps onto initial momenta of $130$ to $250\MeV$. The strength from above $300\MeV$ amounts to $10$ to $30\%$ of the quasi-elastic tail in the window, depending on how hard the momentum distribution falls. The baseline error keeps small given a series of effects. First, the quasi-elastic tail is only $60\%$ of the one-body dip yield as the $\Delta$ part is budgeted separately. Second, even an uncontrolled $30\%$ uncertainty on the high-momentum strength would then move the baseline by only a few percent. Third, the spectator spectrum of the tagged one-body sample measures that high-momentum strength in situ. With this constraint, the extrapolated baseline stays within the $2$--$4\%$ as in Tab.~\ref{tab:req}. The high-momentum sample alone would cause tens-of-percent uncertainty as stated above, resulting in the fact that the high-momentum sample serves as an exploration area.

In addition, over the dip region, the exclusive final-state selection vetoes most of the one-body $\Delta$ strength. On the electromagnetic
side, the $\Delta^+\to p\pi^0$ two-thirds is removed by the $\omega^*$ migration
and $\pi^0$ photon veto, the $n\pi^+$ third by the charged-pion final state. The residual
$\Delta$ then leaves the vector baseline at the $2$--$4\%$ level.
The charged-current baseline is less well known in the same regime, because with the one-body $\Delta$ inside $\sigma_\IA$, its accuracy is set by $\Delta$ electroproduction on the bound neutron and by the $C_5^A$ form factor. Realistically, this is at the few-percent level and it enters the axial-current uncertainty budget alongside the $(A/V)_m$ theory floor. 

One further budget term covers the final-state-interaction asymmetry between the channels. Rescattering conserves the pair isospin. 
The subtraction compares an outgoing $pn$ pair in the electromagnetic channel with an outgoing $nn$ pair in the charged-current channel. The measured $np$ and $pp$ cross sections at the relevant relative momenta differ by $20$ to $45\%$, because the $np$ pair also interacts in the $T=0$ channel that identical nucleons lack~\cite{Arndt2007}. The two-body signal itself is isovector with $T=1$ final states. For those states, the $pn$ and $nn$ interactions are equal by charge independence, and cancels in the subtraction. What cannot cancel is the one-body tail that leaks into the dip window, given that tail carries a $T=0$ component present only in the electromagnetic channel. Folding the few-percent leak with the $20$ to $45\%$ kernel difference gives an uncancelled residue of $0.5$ to $2\%$ of the baseline. It is largest at low $Q^2$, where rescattering is strongest~\cite{BoeglinSargsian2015,CosynSargsian2011}.
We carry this as a budget line. A channel-by-channel distorted-wave comparison in a single framework will replace the estimate. We flag it as a separate dedicated calculation.

\subsection{Mechanism decomposition and V--A connection}
\label{sec:mechreq}

The mechanism decomposition and the V--A connection carry the program beyond a single measured number.
Both rest on the polarized observables, which impose requirements of their own
beyond the unpolarized and charged-current ones.

The mechanism fractions are obtained by inverting the weight matrix of
Eq.~(\ref{eq:weights}). The matrix is well conditioned only because the three mechanisms imprint different patterns across the responses. In particular, the
$\Delta$ contribution changes sign in the tensor-transverse response.

\begin{figure*}[t]
\centering
\includegraphics[width=\textwidth]{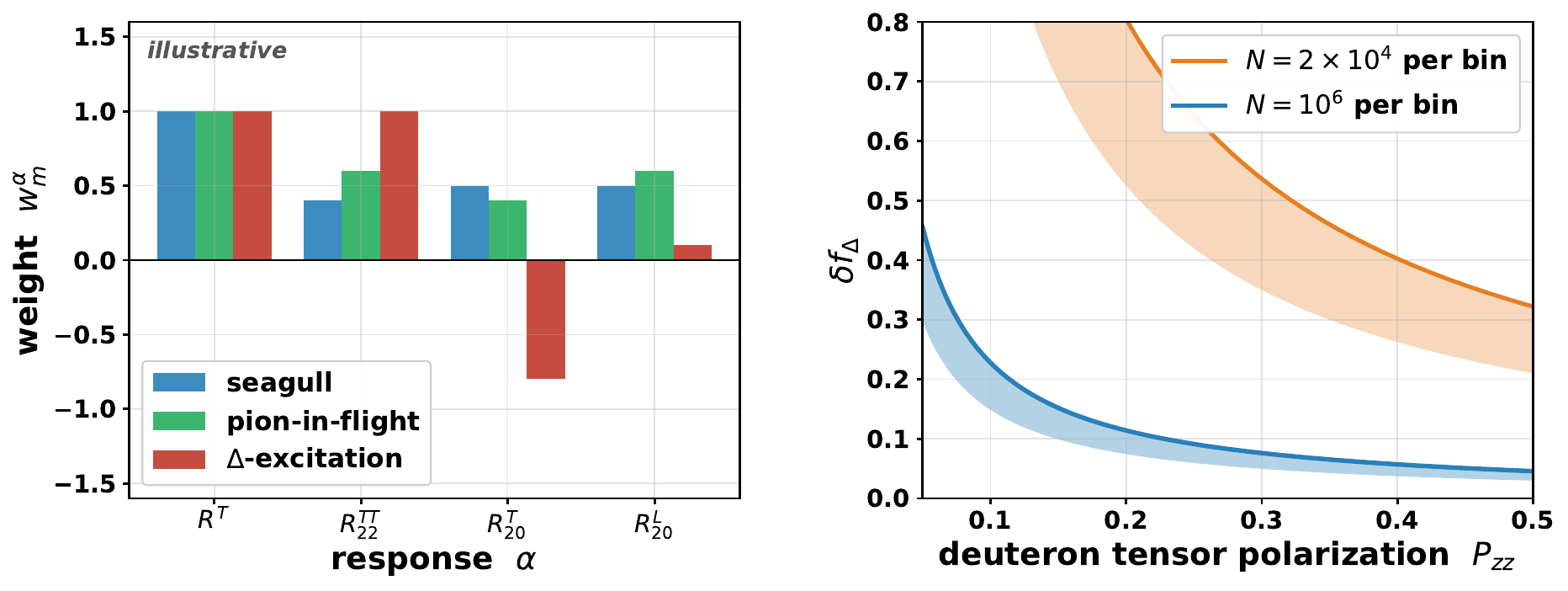}
\caption{Left: illustrative weights $w_m^\alpha$ of the three mechanisms in the
transverse ($R^T$), tensor ($R^{TT}_{22}$), tensor-transverse ($R^T_{20}$), and tensor-longitudinal ($R^L_{20}$)
responses. These are collider-accessible (unpolarized and tensor-polarized) responses. The $R^T$ row has equal weights and serves as the normalization constraint, so the remaining rows overdetermine the two independent fractions. The $\Delta$-excitation weight changes sign in $R^T_{20}$, making the system invertible and providing the distinctive $\Delta$ test. Right: the uncertainty
on the $\Delta$ fraction $\delta f_\Delta$ as a function of the deuteron tensor
polarization $P_{zz}$. The bands include the one-body dilution $(1+V)/V$ for $V=0.15$ to $0.25$. The requirement $\delta f_\Delta=0.1$, needed to separate the leading models met at $P_{zz}\gtrsim0.2$ with the $N=10^6$ sample.}
\label{fig:mech}
\end{figure*}

In Fig.~\ref{fig:mech} left panel, the illustrative weights for each mechanisms are shown. Recall that the $\Delta$ current is the only purely spin-flip mechanism. A spin-flip current enters the tensor-transverse response with the opposite sign relative to the seagull and pion-in-flight currents, making that matrix invertible.

%This argument holds within the current models. Establishing it beyond them calls for a dedicated calculation of the tensor-transverse two-body response. 
%The dip-region tensor-transverse two-body response has not been calculated in any framework. 
The decomposition in this subsection is mainly contingent on the sign structure. However, the model was not built for tensor observables. 
Establishing them calls for a dedicated calculation of the tensor-transverse two-body response. 
The dedicated calculation at dip kinematics is a stated prerequisite for Measurement~III.

Accessing different mechanism responses requires a
tensor-polarized deuteron, a capability not present in any current beam.
The precision on the $\Delta$ fraction follows from counting. Assume a
representative dip dataset of $N\sim\text{few}\times10^{4}$ events per $Q^2$ bin,
reached with under a $\fbi$ at the electromagnetic dip rate. This is deliberately far below the raw $10^{6}$--$10^{7}$ events per bin per $\fbi$ in Sec.~\ref{sec:rates}. The effective sample folds in the exclusive spectator-tagged selection, the signal region, and the share of running taken with a tensor-polarized target. The polarization requirement derived below is therefore conservative. 
If the full inclusive rate could be used, the same $\delta f_\Delta$ would be reached at lower $P_{zz}$. 

The statistical error on the $\Delta$
fraction then scales as $\delta f_\Delta\approx3\times\frac{1+V}{V}\times\frac{1}{P_{zz}\sqrt N}$. This scaling follows from standard error propagation. 
The three comes from inverting the weight matrix $w_m^\alpha$. The factor $(1+V)/V\approx5$ to $8$ is the one-body dilution. The tensor asymmetries are measured on the full dip yield while the fractions live in the two-body excess. The one-body part carries its own D-state tensor response, and that response is subtracted using the same impulse-approximation machinery as the unpolarized baseline, with its uncertainty in the budget. 

For example, at $P_{zz}=0.3$ and $N=2\times10^4$, the calculation gives $\delta f_\Delta\approx0.4$. To separate the Valencia and SuSAv2 mechanism content, which differ by $\Delta f_\Delta\approx0.2$, at two standard deviations requires $\delta f_\Delta\le0.1$. That needs $N\approx10^6$ selected tensor-polarized events per bin at $P_{zz}=0.2$, which corresponds to ten to one hundred inverse femtobarns of tensor-polarized running against the raw electromagnetic rate. The same inversion returns $f_{\rm sea}$ and $f_\pi$ with errors of similar size, separated mainly by their $Q^2$ trends.

In Fig.~\ref{fig:mech} right panel, we show
how the $\Delta$-fraction uncertainty falls with the tensor polarization, including the one-body dilution, and meets the $\delta f_\Delta=0.1$ requirement near $P_{zz}=0.2$ for the $10^6$ event sample.
The tensor and interference responses are extracted from the azimuthal modulation of the cross
section, so they require good azimuthal coverage and reconstruction of
the recoil. In the electromagnetic channel the struck proton is tracked and its azimuth is well measured. The dip-resolution requirement applies unchanged, since a finite $\sigma_\omega$ dilutes the
asymmetry amplitude in proportion to the leakage it causes.

Backgrounds are much less a worry here than in the charged-current channel.
These are polarization asymmetries, so an unpolarized background does not fake the signal but dilutes it. If a fraction $b$ of the events is unpolarized
background, the measured asymmetry is reduced by $(1-b)$ and the statistical error is
inflated by $1/(1-b)$. Keeping this dilution below $10\%$, so that the asymmetry is
not significantly degraded, requires the unpolarized background fraction to satisfy
$b\lesssim0.1$. This is a modest requirement, within the reach of the
electromagnetic event selection. The
limiting systematic is the knowledge of the beam and target polarizations.

The V--A connection turns this decomposition into an axial determination. With the
fractions in hand and the calculable per-mechanism ratios $(A/V)_m$, the axial
excess is predicted from the measured vector excess through
Eq.~(\ref{eq:vaconnection}).
Because the $(A/V)_m$ are all in order of one, their weighted average in Eq.~(\ref{eq:vaconnection}) depends weakly on the fractions. 
The pion mechanism is treated separately in the scan. For the electron beam, the pion-pole part (Sec.~\ref{sec:ltsep})
of its axial current is removed by the lepton mass, so $(A/V)_\pi$ is well below the other mechanisms. 

The export to neutrinos restores exactly that piece through partially conserved axial current (PCAC), as described in Sec.~\ref{sec:impact}.
The prediction is limited by the $(A/V)_m$ theory across the whole range of $\delta f_\Delta$. Its floor is essentially the theory uncertainty. 

The statistics-rich, background-robust electromagnetic program constrains the axial current at the $25$ to $45\%$ level, the quadrature of the measured $\delta V_{\pn}/V_{\pn}=15$ to $31\%$ and the $10$ to $30\%$ theory spread on the $(A/V)_m$. This can be done
with the polarized electron data alone, before the charged-current luminosity is in hand. 

\begin{figure}[t]
\centering
\includegraphics[width=\columnwidth]{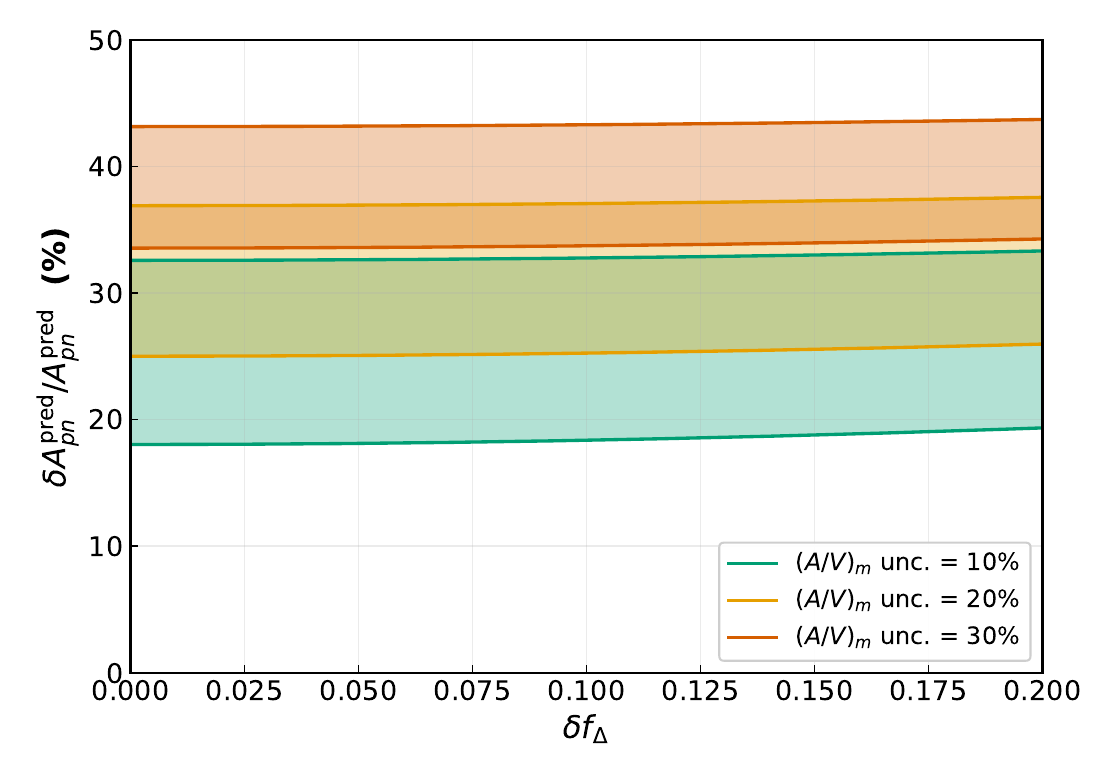}
\caption{Relative uncertainty on the indirect axial prediction
$A_{\pn}^\text{pred}$ as a function of the $\Delta$-fraction uncertainty
$\delta f_\Delta$, for three values of the theory uncertainty on the per-mechanism
$(A/V)_m$ ratios. Each band spans the measured $\delta V_{\pn}/V_{\pn}=15$ to $31\%$ from the baseline control in Fig.~\ref{fig:ia}. With the $(A/V)_m$ all in the order of one, their weighted average depends only weakly on the fractions.}
\label{fig:va}
\end{figure}

In Fig.~\ref{fig:va}, we show the relative uncertainty on the predicted
$A_{\pn}$ as a function of the $\Delta$-fraction uncertainty, for three values of the
$(A/V)_m$ theory error.
The direct charged-current subtraction, when it becomes feasible, will be a test of that prediction.
The two routes share the measured $V_{\pn}$. Now write the predicted ratio of the axial to the vector excess as $k=\sum_m(A/V)_m f_m$ from Eq.~(\ref{eq:vaconnection}). The indirect route then predicts the charged-current excess to be $(1+k)V_{\pn}$. The test compares the direct measurement with this prediction through the difference $D=(V\!+\!A)-(1+k)V_{\pn}$. It vanishes if the prediction holds. Then because the same measured $V_{\pn}$ enters both terms, the shared error enters with weight $(1+k)\approx2$. The variance of $D$ carries $(1+k)^2\,\mathrm{var}(V_{\pn})$. The discriminating power includes this correlation.

Following that note, the direct route carries its own uncertainty. The counted quantity is the charged-current rate in the signal region. Meanwhile, the baseline is computed. The Poisson error enters the excess with the factor $1+V+A$. The residual background discussed in Sec.~\ref{sec:bkg} sits under the signal at about a fifth of it at the favorable end of the background range and then its fluctuations inflate the rate error by $\sqrt{1+B/S}\approx1.1$. At a few hundred $\fbi$, the rate error is then $\approx11\%$ and becoming $\approx15\%$ on the excess. Combining it with the electromagnetic baseline control ($r_{\rm EM}\approx3\%$) and the charged-current baseline control ($r_{\rm CC}\approx4\%$), we have $\delta A_{\pn}\approx0.16$ to $0.18$. At the working expectation $(A/V)\sim1$ to $2$, this is roughly $60$ to $100\%$ of $A_{\pn}$, given that $\delta A_{\pn}\approx0.16$ against $A_{\pn}=0.15$ at $(A/V)=1$ and $\approx0.18$ against $0.30$ at $(A/V)=2$. 

In above calcuation, we use the favorable half of the expected $(A/V)$ range. If the axial strength were at the lower end, $(A/V)\sim0.5$, the same error would be $\approx200\%$. 
Thus, the headline assumes an axial strength at least comparable to the vector one. The systematic floor of the subtraction sits at $20$ to $40\%$ so the counting term falls below the systematics only beyond roughly $1700\fbi$.

At a few hundred inverse femtobarns, this is a $0.9$ to $1.7$ standard deviation statement per bin for an axial strength between one and two times the vector one. At $50\fbi$, it is roughly half of that. Distinguishing those two hypotheses per bin is below one standard deviation. Integrated over the roughly five dip bins the statistical error improves by a factor of about $2.2$. Then the existing statement reaches two to three standard deviations, and the statistics-to-systematics crossover moves from about $1700$ to about $370\fbi$. The integrated measurement carries the first-observation weight while the per-bin numbers govern the shape.

A separate concern is the precision of $r_{\rm CC}$. We just claimed it above but it will actually set requirements on the experiment. The value of $r_{\rm CC}$ follows from the veto chain more than the raw $C_5^A$ spread. The spread of $10$ to $20\%$ on $C_5^A$~\cite{Graczyk2009,Hernandez2010} doubles to $20$ to $40\%$ on the one-body $\Delta$ rate. The $\Delta$ tail carries about $0.4$ of the dip window. Without suppression this would be $8$ to $16\%$. The exclusive selection changes this. The $p\pi^-$ third of the $\Delta^0$ decays leaves charged tracks thus vetoed. The $n\pi^0$ two thirds loses the pion energy and only about a fifth of it migrates into the dip window. That remainder is suppressed by the $\pi^0$ photon veto with efficiency $\varepsilon_{\pi^0}$. For $\varepsilon_{\pi^0}=0.8$ the residual $\Delta$ share of the window is about $2\%$. On the other hand, the induced baseline uncertainty from the $C_5^A$ spread is below one percent, inside the quoted $r_{\rm CC}\approx4\%$. Hence, the veto efficiency for the $15$ to $30\GeV$ boosted photons is a requirement (Tab.~\ref{tab:req}). For reference, at $r_{\rm CC}=8\%$ the systematic floor reaches $0.11$ absolute, about three quarters of the $(A/V)=1$ signal, and the statistics crossover moves from about $1700$ to about $550\fbi$.

One further caveat is specific to the axial sector that the indirect prediction covers only part of the measured excess. The $(A/V)_m$ of Tab.~\ref{tab:mech} are built for the transverse mechanisms while the counted charged-current excess at $\varepsilon\approx1$ also contains the charge and longitudinal responses (Sec.~\ref{sec:ltsep}). 
However, for the vector current, this mismatch is harmless because current conservation forces the vector longitudinal excess below the few-percent level. The axial current obeys no conservation law so its charge and longitudinal responses are unconstrained by that argument. Here, one relevant existing measurement is the nuclear beta decay. In the transitions called first forbidden, the decay proceeds through the time component of the axial current instead of the usual spin operator. Their measured rates exceed one-body calculations, with the axial-charge matrix element enhanced by up to a factor of two. 
An effect of that size in the dip would sit inside our measured excess with no counterpart in the prediction. 
So we carry an explicit axial-longitudinal term in the comparison, to be bounded by a dedicated evaluation of the five weak responses at dip kinematics in the relativistic 2p2h framework in Ref.~\cite{RuizSimo2017}. Until that bound is in hand, an excess of the direct measurement over the transverse prediction would first be attributed to this known source instead of new axial dynamics.

\subsection{Reconstructing the charged-current dip}
\label{sec:ccdip}

Although we touched several charge-current requirements already in the previous sections, it is still worth a dedicated discussion due to the difficulty. Apparently, since the struck nucleons in a charged-current event emerge as neutrons, the dip energy transfer must be reconstructed from forward neutrons in a zero-degree calorimeter. 
The calorimeter measures each neutron's energy from its hadronic shower, with the characteristic $50\%/\sqrt{E}\oplus\sim\!5\%$ resolution set by the large fluctuating shower. The other way is the time-of-flight, which is particularly good for slow neutrons. At $E\approx120\GeV$, the neutron is ultrarelativistic, with $\beta=1-m_N^2/2E^2\approx1-3\times10^{-5}$. So it lags a photon by only $\sim\!3.6$ ps over the $\sim\!35$ m flight. Reaching the calorimeter's few-percent energy resolution in this way would demand picosecond timing. The forward-neutron energy is likely calorimetric, and time-of-flight enters this program for background rejection. The two signal neutrons, both essentially at $c$, are separated spatially by the calorimeter granularity.

The per-neutron resolution could be adequate. The rest-frame neutron energy is $E^*=\gamma(E-\beta p_z)$, which splits into two pieces, $E^*=\gamma(E-p_z)+p_z/2\gamma$. Differentiating at fixed angle gives $dE^*/dE=1/2\gamma+\gamma(\theta^2/2-m_N^2/2E^2)$. 
At the situation of $\gamma=100$, $E=120\GeV$, $\theta\approx3$~mrad, the two contributions largely cancel, and then the net sensitivity is $\approx2\times10^{-3}$. 

Then the $\sim\!8\GeV$ shower error leaves a floor of about $25\MeV$ per neutron. The naive boost factor $\gamma(1+\beta)$ acting on $E-p_z$ alone would double this floor, and it is the $p_z/2\gamma$ piece that removes the excess. The angle enters linearly about the typical neutron angle, $dE^*/d\theta\approx\gamma E\theta_0$ with $\theta_0\approx3$~mrad for the dip transverse momenta, giving about $36\MeV$ per milliradian. Adding the two in
quadrature,
$\sigma_{\omega^*}\approx\sqrt{25^2+(36\,\sigma_\theta[\text{mrad}])^2}\MeV$ per neutron. The reconstructed $\omega^*$ combines both signal neutrons, so the per-neutron budget is $60/\sqrt{2}\approx42\MeV$. Note that this estimate holds at the reference energy of 120 GeV.

The actual signal neutrons are more complicated. The pair decay spreads them over $E\approx50$ to $140\GeV$, where the energy floor grows to $30$ to $47\MeV$ per neutron and the angle term is evaluated at the accepted angles of $3$ to $5$~mrad. We evaluate the budget over the accepted population event by event, as shown in Fig.~\ref{fig:popbudget},
with $\omega^*$ sampled over the signal window. At the baseline resolutions, the median two-neutron $\sigma_{\omega^*}$ is $47$, $64$, and $90\MeV$ in the three reference bins. There are two ways to recover the budget. A tighter angular resolution of $\sigma_\theta\approx0.2$ to $0.3$~mrad restores it in the low and central bins. A better option is a constrained refit using $E_{n1}+E_{n2}=E_d$ and transverse-momentum balance, meaning two constraints on four measured quantities. It could recover roughly a factor $\sqrt2$. The refit medians are $33$, $45$, and $64\MeV$, restoring the budget at the baseline resolutions. 

\begin{figure}[t]
\centering
\includegraphics[width=\columnwidth]{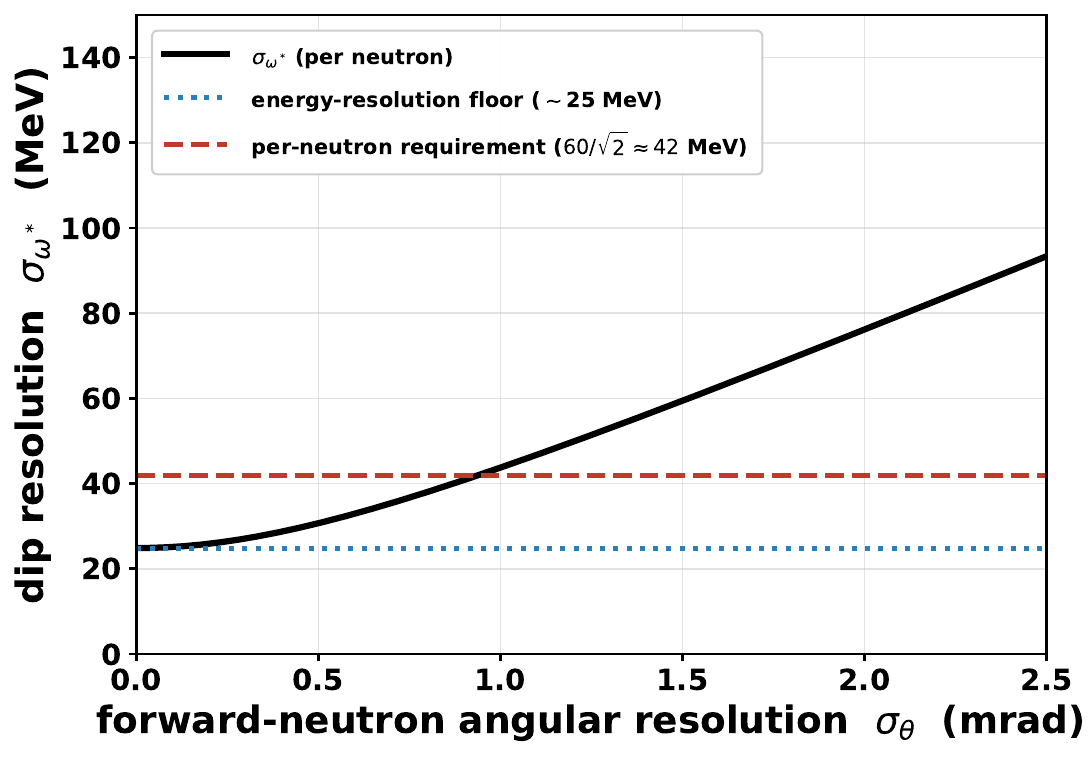}
\caption{Charged-current dip resolution per forward neutron as a function of the
neutron angular resolution $\sigma_\theta$. The solid curve is $\sigma_{\omega^*}$.
The dotted line is the energy-resolution floor ($\sim\!25\MeV$), and the dashed line
is the per-neutron requirement of $60/\sqrt2\approx42\MeV$. }
\label{fig:neutron}
\end{figure}

In Fig.~\ref{fig:neutron}, we show the fiducial per-neutron resolution as a function of the angular resolution, rising from the $25\MeV$ energy floor toward the per-neutron requirement line. The curve crosses the $42\MeV$ per-neutron budget near $\sigma_\theta\approx0.9$~mrad. At $35$~m, this corresponds to a transverse position resolution of about $3$~cm, which centimeter-scale calorimeter granularity provides. The per-neutron requirement remains satisfiable. 

On the other hand, the missing transverse momentum of the signal is reconstructed from the same two neutrons. The per-neutron resolution combines the angle and energy terms, $\delta p_T=\sqrt{(E\,\sigma_\theta)^2+(\theta_0\,\delta E)^2}$, with the energy term contributing $24\MeV$ at $\theta_0\approx3$~mrad. The background rejection of Sec.~\ref{sec:bkg} assumes $80\MeV$ for the two-neutron combination, which requires $\sigma_\theta\lesssim0.43$~mrad. At that value the per-neutron $\omega^*$ resolution is $29\MeV$ and the two-neutron combination is $42\MeV$, which is inside the budget and meanwhile, the position resolution at $35$~m is $1.5$~cm. Therefore, we adopt $\sigma_\theta\lesssim0.43$~mrad for the missing-transverse-momentum reconstruction, with the population $\omega^*$ budget met through the constrained refit or the tighter $\sigma_\theta\approx0.2$ to $0.3$~mrad.

Above said, the main limitation is the
separation of the two forward neutrons of the 2p2h final state. Both carry an order of $100\GeV$ into the small forward cone. Their separation at the calorimeter reaches up to $\sim\!20$ cm for the largest opening angles but shrinks to the centimeter
granularity scale for the smallest. Current single-arm calorimeters cannot resolve the
small-opening-angle pairs. A high-granularity, large-acceptance forward-neutron calorimeter is the specific capability the charged-current measurement demands. One caution applies. At tight apertures the accepted pairs are the small-opening ones, with a median transverse separation of a few cm at the calorimeter. Geometric acceptance and two-cluster resolvability are anticorrelated so the product in the yield chain is conservative only once the cluster-separation requirement of Tab.~\ref{tab:req} is met.

{\color{blue}\begin{figure}[t]
\centering
\includegraphics[width=\columnwidth]{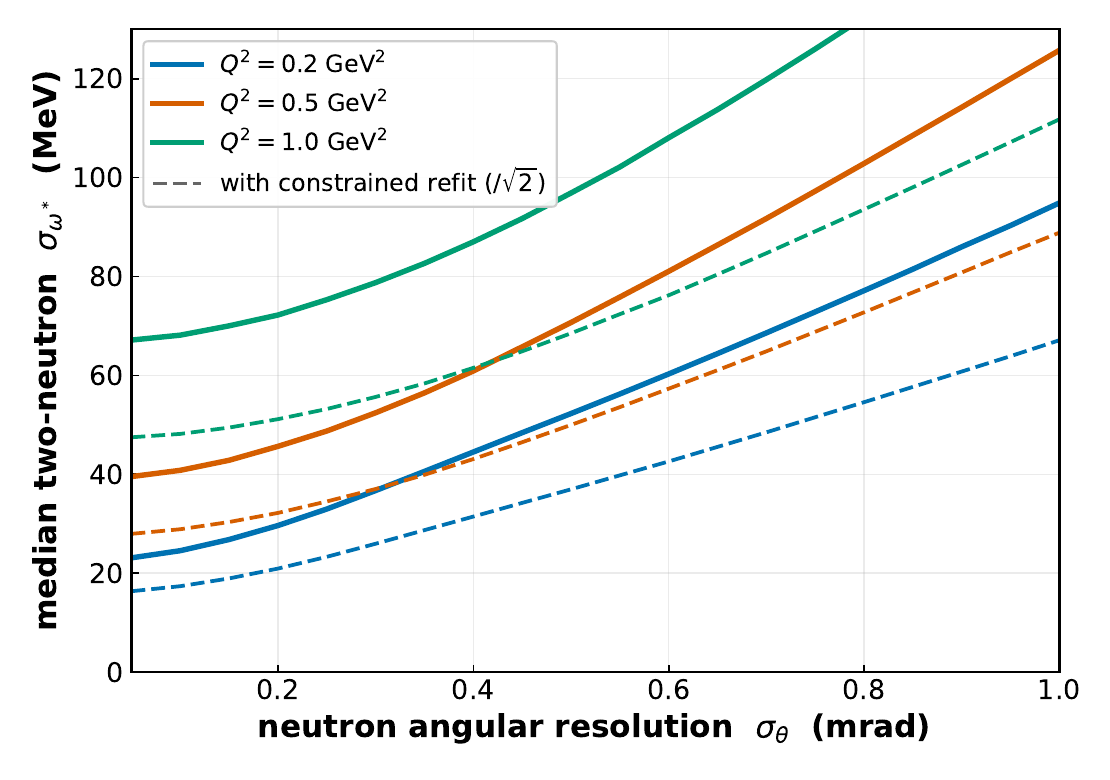}
\caption{Median two-neutron $\sigma_{\omega^*}$ over the accepted population versus the neutron angular resolution, per $Q^2$ bin at the maximum-rigidity configuration. Dashed curves include the constrained refit. The two-neutron requirement is $60\MeV$.}
\label{fig:popbudget}
\end{figure}}

\subsection{Charged-current backgrounds}
\label{sec:bkg}

At this point, we went through a few signal handling tricks. However, signal reconstruction is only part of the charged-current problem that needs to be separated from copious backgrounds. The charged-current signature without a scattered electron can be mimicked by photoproduction, in which the electron radiates a near-real photon and escapes down the beam pipe. The photoproduction rate exceeds the signal by eight orders of magnitude, obtained by looking at the ratio of the effective photoproduction cross section built below from the photon flux and the total $\gamma p$ cross section~\cite{PDG2024} to the $9\fb$ signal scale, so the rejection must be
correspondingly large. 
In elastic CC, the final state is a single forward neutron, and background suppression was found to be the controlling difficulty~\cite{YangKumar}. Here the signal demands two forward neutrons in the zero-degree acceptance together with the missing transverse momentum and the dip kinematics. Each added a handle thus brings the required $\sim\!10^{8}$ within reach.

The individual factors are order-of-magnitude estimates with the following basis. They need to be confirmed by a full simulation. The starting rate is the effective
$ep$ photoproduction cross section. The quasi-real photon flux ($\sim\!1\%$ per electron, Weizs\"acker--Williams) times the $\gamma p$ cross section of $\sim\!0.15\,$mb gives $\sim\!1.5\,\mu$b per nucleon~\cite{PDG2024}. On a deuteron beam this roughly doubles, $\sigma_{\gamma d}\approx2\sigma_{\gamma p}$, giving $\sim\!3\times10^{9}\fb$, about $10^{8}$ times the
$9\fb$ signal. The missing-$p_T$ cut gives the largest single suppression. A quasi-real photon carries little transverse momentum, peaked at the MeV scale, so a genuine
photoproduction event is $p_T$-balanced. 

On the other hand, the signal has the neutrino carrying
off $p_T$ an order of the momentum transfer, several hundred $\MeV$. Requiring the reconstructed missing $p_T$ above that scale keeps photoproduction only through the
tail of the detector $p_T$-response. 
In this final state, the missing transverse momentum is carried by the two forward neutrons, so its resolution is set by their angular and energy resolutions, $\approx\!80\MeV$ per component at the $\sigma_\theta\lesssim0.43$~mrad shown in Sec.~\ref{sec:ccdip}. 
For a background event the true missing transverse momentum is zero and both measured components fluctuate independently. The chance that the magnitude exceeds a cut at $k$ times the per-component resolution is then $e^{-k^2/2}$. The Rayleigh tail falls slower than the Gaussian tail.
We place the cut at $\approx\!0.4\GeV$, where the Gaussian-core Rayleigh tail is $4\times10^{-6}$. 

Large missing transverse momentum in photoproduction is dominated by whole particles escaping through acceptance gaps, and those tails are set by the gap solid angles rather than by resolution. So we present $10^{-5}$ as a target to be demonstrated with data-driven control samples. The residual-control strategy below is designed around this measurement. The cut sits well below the signal scale of $\sqrt{Q^2}\approx0.7\GeV$.

Next is regarding the neutron rigidity for $\omega^*$ reconstruction. On an ion beam the first forward neutron is not rare. Photoproduction on the proton dissociates the deuteron, and the untouched neutron continues at beam rigidity into the zero-degree acceptance. We take a probability of $0.3$ to $0.5$ for this spectator branch, set by the dissociation probability and the calorimeter efficiency~\cite{PaulZDC2024,ZhangZDC2025}. 
Then the second must come from the produced hadrons. Here an energetic forward neutron is rare, at the $10^{-2}$ level on a free proton. 
In addition, the dip kinematic selection contributes the next $\sim\!10^{-2}$, given the fact that the reconstructed $\omega^*$ of the fake pairs populates the full spectrum and only a small fraction falls in the narrow dip window.

The other factor comes from energy conservation. For signal the neutrino carries negligible laboratory energy, so the two neutrons must sum to the full ion energy, $E_{n1}+E_{n2}\approx E_d$ to MeV accuracy. With the $7\%$ calorimetry the summed peak has $\sigma\approx12\GeV$. The dangerous background topology pairs a beam-rigidity spectator with a produced neutron of much lower energy, and thus its sum populates a broad continuum well below $E_d$. A cut at two sigma retains essentially all signal and suppresses this topology by a further factor of $0.1$ to $0.3$. 

\begin{figure}[t]
\centering
\includegraphics[width=\columnwidth]{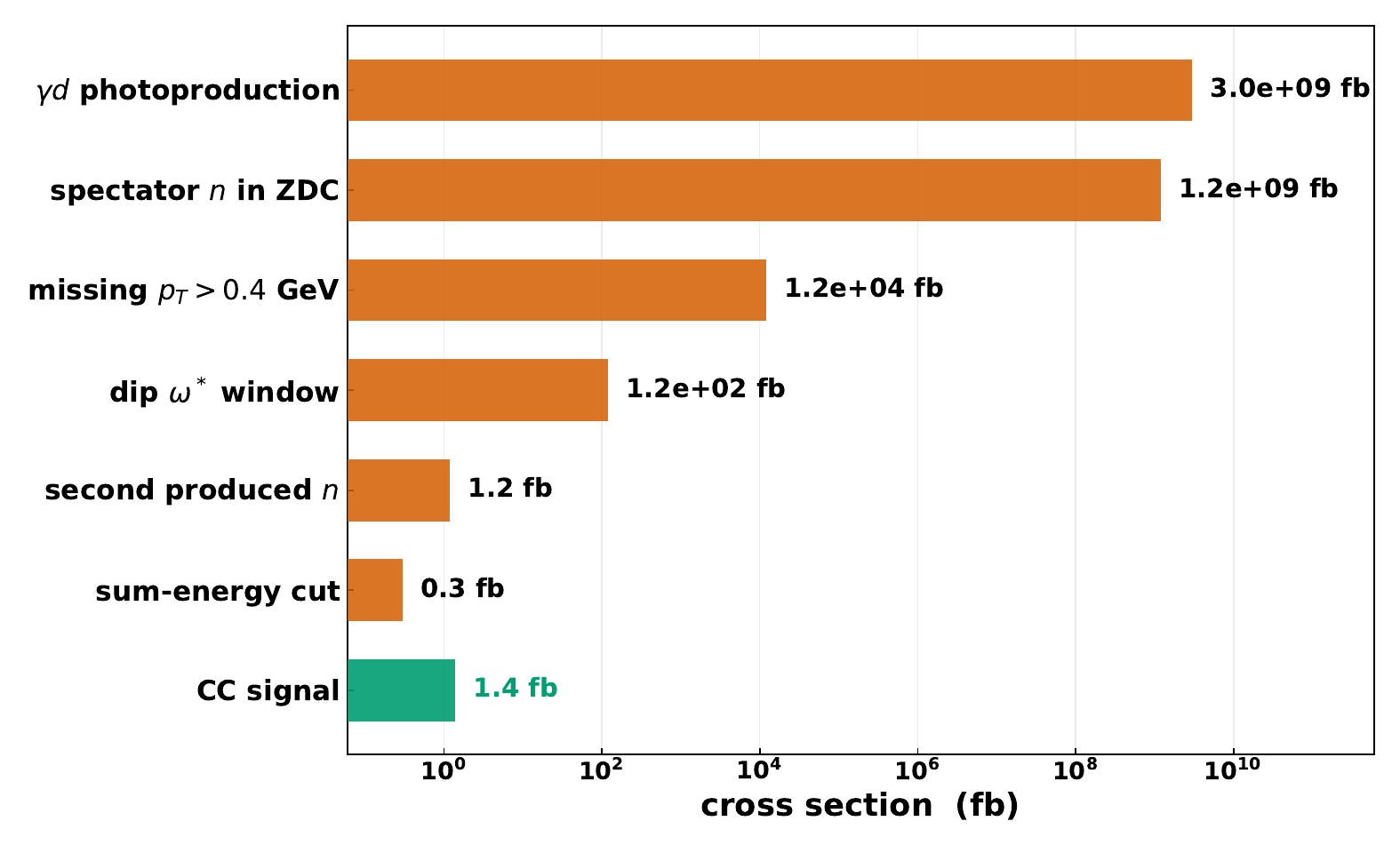}
\caption{Ion-beam rejection chain for quasi-real photoproduction on the deuteron. The first stage uses $\sigma_{\gamma d}\approx2\sigma_{\gamma p}$. The first forward neutron comes from the spectator branch. The missing-$p_T$ factor is a target to be demonstrated with control samples. The green bar marks the selected charged-current signal.}
\label{fig:bkg}
\end{figure}

Putting the chain together, the product of the spectator branch, the missing-$p_T$ target, the dip selection, the second-neutron factor, and the sum-energy requirement brings the residual to $\approx\!0.3\fb$. On the same selection footing the signal is $\approx\!1.4\fb$ over the dip range (Sec.~\ref{sec:rates}). Therefore, the matched background-to-signal ratio is $B/S\approx0.2$ to $0.5$.
The rejection formed
from the individual cuts is shown in Fig.~\ref{fig:bkg}.

It is also important to note that the factors are not fully independent. A photoproduction event acquires a forward neutron by losing particles down the beam pipe while the same loss fakes missing transverse momentum. So the neutron-tag and missing-$p_T$ suppressions are positively correlated. The error estimate is a order of magnitude evaluation, instead of a complete budget.

Reducing the background is not the end point. The observable is the two-body excess, which is $15$ to $25\%$ of the charged-current rate. The residual background sitting at the level of the counted excess on the matched selection must be determined from data to roughly ten percent of itself before the excess can be measured. The $\omega^*$ sidebands outside the dip region, the non-colliding bunches, and reversed-cut control samples all measure the residual in place with large statistics. The quoted signal yields include the geometric two-neutron acceptance and those efficiencies scale the luminosity requirement in Sec.~\ref{sec:reqsummary} accordingly.

Lastly, beam-gas interactions, which do not point to the luminous region, are measured and subtracted
using non-colliding bunches. It is a standard technique at the HERA
experiments~\cite{H1detector}.
A reliable background estimate requires a
simulation that folds in the resolution-induced migration of these processes into the dip. 

\subsection{Summary of requirements}
\label{sec:reqsummary}

The electromagnetic program is statistics-rich and needs only modest integrated luminosity, but it requires tensor polarization of the deuteron at the collider. The charged-current program is counting-limited and requires an order of a few hundred $\fbi$ to isolate $A_{\pn}$.
We summarize the requirements in Tab.~\ref{tab:req}. They fall into three
categories. Several are already met by existing experiment designs such as the Electron-Ion Collider (EIC) including the spectator-proton
momentum resolution ($\delta p/p\sim10^{-3}$) and the electromagnetic dip energy-transfer
reconstruction through the spectator tag. The few-percent baseline control at low
spectator momentum is met not by detector design but by analysis, given the far-forward
acceptance to select low spectator momentum (with caveat of Sec.~\ref{sec:iabase}). Others are design-compatible but not yet in the baseline including the
few-hundred-$\fbi$ integrated luminosity and the high-granularity forward-neutron
calorimetry needed to resolve the two signal neutrons. The genuinely new capability is polarized deuteron beams, tensor-polarization for the collider mechanism decomposition. These specifications lie outside the present
polarized species ($e$, $p$, $^3$He) in experiments like EIC. Integration of all the components defines the instrument needs for the program.

\begin{table*}[t]
\caption{Summary of experimental requirements.
Each defines a capability a future high-luminosity polarized light-ion collider
could provide.}
\label{tab:req}
\begin{ruledtabular}
\begin{tabular}{ll}
Requirement & Value \\
\colrule
Dip energy-transfer resolution & $\sigma_{\omega^*}\lesssim60\MeV$ \\
$L/T$ separation & not feasible; $V_{\pn}$ survives \\
Spectator tag (EM, $^3$He CC) & $\delta p/p\sim10^{-3}$ \\
CC dip, $\omega^*$ population budget & $\sigma_\theta\approx0.2$ to $0.3$ mrad or constrained refit \\
CC dip, limiting factor & two-neutron forward separation \\
CC background rejection & ion-beam chain in Fig.~\ref{fig:bkg}, $B/S\approx0.2$ to $0.5$ \\
CC residual-background control & data-driven, $\approx$10\% of itself \\
Polarized-asymmetry background & dilution $b\lesssim0.1$ \\
Scattered-electron acceptance & $|\eta|$ reach, favors $E_e\lesssim10\GeV$ \\
Scattered-electron angle & $\sigma_\theta$ of a few mrad for $Q^2$ binning \\
IA-baseline control (EM) & $r\sim2$--$4\%$ (anchored at low $p_s$, extrapolated) \\
CC-baseline uncertainty from $C_5^A$ & few \%, feeds the axial budget \\
Polarization, electron beam & $P_e$ (fixed-target program) \\
Polarization, deuteron & $P_{zz}$ collider, $P_z$ fixed target \\
Mechanism decomposition & $P_{zz}\gtrsim0.2$ with $N\approx10^6$ per bin ($10$--$100\fbi$ tensor polarized) \\
V--A connection (indirect axial) & $(A/V)_m$ theory; floor $20$--$30\%$ \\
Direct axial precision (CC$-$EM) & $\approx60$--$100\%$ at few$\,\times100\fbi$ ($A\gtrsim V$), floor $20$--$40\%$ \\
Luminosity (CC/axial) & $\gtrsim$ few$\,\times100\fbi$ (before selection efficiency) \\
ZDC neutral aperture & $\theta_{\max}\gtrsim5.5$ mrad for two-neutron acceptance \\
CC ion energy & maximum deuteron rigidity ($\gamma\approx146$) \\
Two-neutron acceptance & $\varepsilon_{2n}(Q^2)$ in Fig.~\ref{fig:acc2n} \\
ZDC two-cluster separation & few cm at $35$ m, anticorrelated with $\varepsilon_{2n}$ \\
ZDC per-neutron detection & efficiency enters squared in charged-current yields \\
CC $Q^2$ assignment & hadronic $Q^2_{\rm rec}$, migration of Tab.~\ref{tab:migration} folded \\
$\pi^0$ photon veto (ZDC EM) & $\varepsilon_{\pi^0}\gtrsim0.8$ for $15$--$30\GeV$ photons \\
Sum-energy selection & $|E_{n1}{+}E_{n2}-E_d|\lesssim2\sigma$, $\sigma\approx12\GeV$ \\
\end{tabular}
\end{ruledtabular}
\end{table*}

% =============================================================================
\section{Implications for neutrino experiments}
\label{sec:impact}
% =============================================================================

The per-pair currents measured on the light nuclei feed the heavy targets of the neutrino oscillation experiments through pair counting. To the extent that the two-body current is a property of the pair and the medium modifications are subleading, the 2p2h cross section on a nucleus of $Z$ protons and $N$ neutrons scales as
\begin{equation}
\sigma_\text{2p2h}(A)\approx N^\text{SR}_{\pn}(V\!+\!A)_{\pn}\sigma_\text{CC}^\IA
+ N^\text{SR}_{\pp}(V\!+\!A)_{\pp}\sigma_\text{CC}^\IA + \cdots.
\label{eq:scaling}
\end{equation}
Here $N^\text{SR}_{\pn}$ and $N^\text{SR}_{\pp}$ count the pairs at short relative distance. Meson exchange is short-ranged thus only close pairs contribute. Their number grows roughly with $A$ rather than with the total pair combinatorics. 

The absolute counts must come from a many-body calculation, with the deuteron fixing the normalization by construction. The combinatorial numbers $N_{\pn}=ZN$ and $N_{\pp}=Z(Z\!-\!1)/2$ overcount the close pairs. Being used as absolute weights, they would let the two-body rate grow with $A^2$ and put argon an order of magnitude above the observed tens-of-percent scale. Therefore, we use the combinatorics only for the relative $\pn$ to $\pp$ weight. For argon ($Z=18$, $N=22$) that weight is $396$ to $153$, and for oxygen $64$ to $28$. Measured short-range pairs are $np$-dominated through tensor correlations~\cite{SRC}, so the combinatorial weight overstates the $\pp$ share. The bias inflates the like-pair term. The like-pair channel in the $^3$He measurement is needed even though the $pn$ channel dominates.

Nonetheless, in Eq.~(\ref{eq:scaling}), it is written for the charge-lowering current
($W^-$, $p\to n$) probed by our $e^-$ beams (like-pair $pp$) while a
neutrino beam carries the charge-raising current ($W^+$, $n\to p$, like-pair
$nn$). Here note that $N_{nn}=N(N\!-\!1)/2$ ($=231$ for argon, larger than
$N_{pp}$). The measured charged-current $(V\!+\!A)_{\pp}$ maps to the required $(V\!+\!A)_{nn}$ by charge symmetry of
the strong interaction, an assumption at the percent level for these light systems~\cite{MillerNefkens1990,AV18}.

%There is another point in Eq.~(\ref{eq:scaling}). It is written there for the squared strengths alone. At neutrino kinematics, the vector--axial interference enters with full weight (Sec.~\ref{sec:method}) and that term is part of the extrapolation budget. With both squared strengths measured, the unknown is the relative phase. The bound is in the order of the signal at $(A/V)\sim1$. So the generator mapping retains an order-one model term.

%One further piece must be restored by modeling. \revb{As noted in Sec.~\ref{sec:ltsep}, the pion-pole part of the axial current vanishes for our $e^-\to\nu_e$ probe, so the measured axial excess excludes it, while a $\nu_\mu$ beam retains its $m_\mu$-weighted contribution. That piece is added back through the partially conserved axial current (PCAC) relation that ties the pion pole to the rest of the axial current when mapping to neutrino generators.}
%The extrapolation assumes the per-pair current is universal. Medium effects, Pauli blocking, and genuine three-body currents enter at the tens-of-percent level on heavy targets and must be computed. Even so, the measurement replaces an unconstrained input of the momentum dependence of the axial two-body current with a data-driven one.

% =============================================================================
\section{Discussion}
\label{sec:discussion}
% =============================================================================

The requirements of Sec.~\ref{sec:req} are stated without directly referring to any operating facility, but the natural question is which existing or planned machine comes close. The method needs a polarized electron--light-ion collider with a
capable far-forward detector suite. Naturally, EIC
is such a facility in prospect~\cite{EICYellow}. Hence, we set our requirements against its capabilities, without asserting that any specific measurement is assured.

The electromagnetic, vector core of the program maps naturally onto the EIC's first detector. The baseline design already provides polarized electron, proton, and $^3$He beams, beam energies that place the dip in the acceptance, and the far-forward suite of Roman Pots, off-momentum detectors, the B0 spectrometer, and the zero-degree calorimeter~\cite{EICYellow,PaulZDC2024,ZhangZDC2025}.

This suite supplies the spectator tagging central to the Fermi-motion correction. The spectator protons of this program sit far off beam rigidity, at around half rigidity on a deuteron beam and at around two thirds on a $^3$He beam. Therefore, they are off-momentum and B0 measurements instead of Roman-Pot measurements. Also the $\delta p/p\sim10^{-3}$ requirement must be delivered by those devices. The spectator deuteron from $^3$He sits at $1.33$ times beam rigidity, a region the present far-forward layouts do not map, so we flag it as an open acceptance requirement. In Tab.~\ref{tab:ffmap}
we collect the rigidity fractions, angles, and device assignments per event class. On this basis the statistics-rich unpolarized measurement of $V_{\pn}$, with the $pp$ extension on $^3$He, is the part of the program most possible within reach of the initial EIC.

{\color{blue}\begin{table*}[t]
\caption{Far-forward destination of each particle class. A magnet bends a particle according to its momentum per unit charge, the rigidity $p/Z$. The ratio $x_L$ of the particle rigidity to the beam rigidity is what we care. Angles are quoted at $\gamma=100$ and
scale as $1/\gamma$. The $nn$ pair carries no charge so sorted by angle alone.}
\label{tab:ffmap}
\begin{ruledtabular}
\begin{tabular}{llll}
Particle & $x_L$ & Angle & Device \\
\colrule
$d$-beam spectator $p$ & $0.50$ & $\lesssim1$ mrad & off-momentum / B0 \\
$^3$He-beam spectator $p$ & $0.67$ & $\lesssim1$ mrad & off-momentum / B0 \\
$^3$He-beam spectator $d$ & $1.33$ & $\lesssim1$ mrad & unmapped, flagged \\
EM struck $p$ & $\approx0.6$ & $4.8$--$10.6$ mrad & B0, low $Q^2$ near the gap \\
SRC-tag $p$ ($p_s\gtrsim300\MeV$) & $0.5\pm0.15$ & spread & spans optics holes \\
CC $nn$ pair & no charge & $0$--$9$ mrad & ZDC, $\varepsilon_{2n}$ in Fig.~\ref{fig:acc2n} \\
\end{tabular}
\end{ruledtabular}
\end{table*}}

It is interesting to note that the electromagnetic and charged-current channels may prefer opposite ion energies. The electromagnetic struck proton needs the lower rigidity for B0 coverage while the charged-current two-neutron acceptance needs the maximum rigidity. Thus the run plan alternates the two configurations. The electromagnetic program takes its luminosity at $\gamma\approx100$ and the charged-current program at $\gamma\approx146$. The excess in Eq.~(\ref{eq:excess}) is acceptance corrected per configuration with a simulation systematic quoted in the budget.

Three requirements reach beyond that EIC baseline design. Each maps onto a concrete design
choice.
The first is the forward-neutron reconstruction. The two signal neutrons carry energy at the order of $120\GeV$ and arrive at a zero-degree calorimeter roughly $35$ m downstream, separated by down to the centimeter granularity scale and up to about $20$ cm for the largest opening angles. Resolving them as two clusters and reconstructing each neutron calls for transverse segmentation at the centimeter scale. In addition, fast timing helps reject out-of-time background. The single-arm zero-degree calorimeter of the first detector is not built for two-neutron imaging. A second EIC
detector's far-forward region could be optimized for this topology with
high-granularity, fast-timing neutron-detection technologies developed for other experiments~\cite{Agarwal2023,Gwon2023,oWbLS2026}, and the zero-degree calorimetry being developed for the EIC itself~\cite{PaulZDC2024,ZhangZDC2025}.

The second is deuteron polarization. At the collider, the request is a tensor-polarized deuteron beam with $P_{zz}\gtrsim0.2$, needed by the mechanism decomposition. Deuteron beams of any polarization sit outside the present program, so this requires development of the polarized source and spin transport through the accelerator. Once again, the vector-polarized responses ($R^{T'}_{\rm EM}$, $R^{TL'}$) are fixed-target observables (Sec.~\ref{sec:helicity}).

The third is luminosity. The charged-current axial reach requires an order of a few hundred $\fbi$. At a delivered luminosity in the order of $10^{34}\,\text{cm}^{-2}\text{s}^{-1}$,
this will be a multi-year, high-luminosity-era program, beyond the first-years projection.

None of the requirements is excluded completely in principle. We set them out in this detail so that the EIC's second detector, luminosity and polarization upgrades can weigh a concrete physics return, namely a data-driven handle on the axial two-body current.

% =============================================================================
\section{Summary}
\label{sec:summary}
% =============================================================================

Combining electromagnetic and charged-current lepton scattering on the same polarized light nucleus isolates the axial-sensitive two-body current as a function of momentum transfer. Such a measurement is missing from the nuclear input to long-baseline neutrino experiments. The axial two-body current is unconstrained today beyond $Q^2=0$.
Performed across the proton, deuteron, $^3$He, and $^6$Li, the method separates the $pn$ and $pp$ channels. Through the spin-dependent responses, it also decomposes
the two-body current into its seagull, pion-in-flight, and $\Delta$-excitation contributions. 

Currently, the leading models on the market disagree on the decomposition. The
tensor-transverse sign change serves as a distinctive test for $\Delta$ dominance, pending the dedicated calculation of the tensor-transverse two-body response flagged in Sec.~\ref{sec:mechreq}. Three further calculations remain stated prerequisites, including the axial-longitudinal bound, the channel-level final-state-interaction comparison, and the MEC-weighted acceptance variant.
That same decomposition predicts the axial current from the electromagnetic data alone, so that the difficult charged-current measurement becomes a direct test of an indirect, statistics-rich prediction.

We have presented the observables and their expected cross sections and yields. We have also quantified what an apparatus must achieve in order to measure them. The electromagnetic part of the program is demanding mainly in its energy-transfer resolution, and in the control of the impulse-approximation baseline. Both requirements are at levels that are stringent but not extreme.
The charged-current, axial part is harder. It is best performed on $^3$He, where a proton spectator can be tagged, and its limiting requirement is the separation of
the two forward neutrons of the signal final state together with the data-driven control of the residual background. This is a capability not provided by
present detectors. 

For an axial strength at least comparable to the vector one, a first direct measurement reaches the axial two-body current at the sixty to one hundred percent level with a few hundred inverse femtobarns, a consistency test of the electromagnetic prediction. These requirements exceed current plans, but they are not fundamentally impossible. Therefore, a clear target for a future high-luminosity polarized light-ion collider can be defined.

\begin{acknowledgments}
In accordance with the APS policy on the use of artificial-intelligence tools, we
disclose that the large language model Claude (Opus~4.6 and 4.8, Anthropic) was used in preparing this work in the following ways: generate the diagrams in
Fig.~\ref{fig:feynman}, debug the analysis and figure-generation code, and
polish the grammar and language of the manuscript. The tool was not used to originate the scientific ideas, reasoning and conclusions. The authors directed, reviewed, and verified all content and take responsibility for the correctness of the results.
\end{acknowledgments}

\appendix
% =============================================================================
\section{The $pp$-pair program}
\label{app:pp}
% =============================================================================

The $pp$-pair currents are reached by adding $^3$He and, for the tensor responses,
$^6$Li, at the cost of additional model dependence that keeps them out of the main
line of the argument. The $^3$He electromagnetic excess contains both $pn$ and $pp$
contributions, and the $pp$ vector current is obtained by subtracting the deuteron
$pn$ result scaled by the ratio of pair densities,
\begin{equation}
V_{\pp}(Q^2)=\frac{X(\he)-f_{\pn}(Q^2)\,V_{\pn}(Q^2)}{f_{\pp}(Q^2)},
\label{eq:vpp}
\end{equation}
the solution of Eq.~(\ref{eq:isospin}) for $V_{\pp}$, with
$X(\he)=[\sigma_\text{EM}(\he)-\sigma_\IA(\he)]/\sigma_\IA(\he)$ and the same
pair-density factors $f_{\pn},f_{\pp}$ computed from the $A=3$ wave function. 
Because $V_{\pp}$ is
the difference of two large quantities, its relative uncertainty is correspondingly
amplified. In addition, it inherits the model dependence on both the $^3$He ground state and
the density-scaling factor. The scaling also assumes the per-pair excess is the same in the deuteron and in $^3$He. In Sec.~\ref{sec:cc}, we note that the $^3$He $pn$ pair is denser and the same mismatch enters here, getting amplified into $V_{\pp}$ by the factor $f_{\pn}/f_{\pp}\approx2$. 

Moreover, polarized $^3$He
functions as an effective polarized neutron, providing the polarized-neutron
spin observables that enter the $pn$ decomposition. These measurements complete the
isospin decomposition and supply the $pp$ input to the heavy-nucleus scaling shown in
Eq.~(\ref{eq:scaling}).

Furthermore, the same subtraction in the charged-current channel yields $(V\!+\!A)_{\pp}-V_{\pp}$, which retains uncancelled vector strength (Sec.~\ref{sec:method}) and is not used as a pure axial observable.
The $pp$ tensor responses, which require a spin-1 target, would call
for tensor-polarized $^6$Li beams that do not presently exist. This part is
exploratory such that $^6$Li is dominantly an $\alpha+d$ cluster, so its tensor polarization is
carried by the deuteron-like $pn$ cluster while the $pp$ pairs sit in the spin-0
$\alpha$ core and are largely unpolarized. The residual $pp$-pair polarization must be
estimated from the $A=6$ wave function before any quantitative $pp$ tensor claim.

% =============================================================================

\end{document}